\documentclass[prd,byrevtex,preprint
,nofootinbib]{revtex4}
\usepackage{amsmath,amsxtra,latexsym,amssymb}
\usepackage[pdftex]{graphicx}
\usepackage{float}
\usepackage{color}
\usepackage{diagbox}
\newcommand{\pa}{\partial}

\begin{document}
\title{Probability of Boundary Condition in Quantum Cosmology}
\author{Hiroshi Suenobu}
\email{Suenobu.Hiroshi@ab.MitusbishiElectric.co.jp}
\altaffiliation[present address: ]{Mitusbishi electric Corporation
  Information Technology R\&D Center, 5-1-1, Ofuna, Kamakura 247-8602, Japan}
\author{Yasusada Nambu}
\email{nambu@gravity.phys.nagoya-u.ac.jp}
\affiliation{Department of Physics, Graduate School of Science, Nagoya 
University, Chikusa, Nagoya 464-8602, Japan}

\date{January 20, 2017}

\begin{abstract}
    One of the main interest in quantum cosmology is to determine
    boundary conditions for the wave function of the universe which
    can predict observational data of our universe. For this purpose,
    we solve the Wheeler-DeWitt equation for a closed universe
      with a scalar field numerically and evaluate probabilities for
    boundary conditions of the wave function of the universe. To
    impose boundary conditions of the wave function, we use
    exact solutions of the Wheeler-DeWitt equation with a constant
    scalar field potential. These exact solutions include wave
    functions with well known boundary condition proposals, the
    no-boundary proposal and the tunneling proposal. We specify the
    exact solutions by introducing two real parameters to
    discriminate boundary conditions, and obtain the probability for
    these parameters under the requirement of sufficient
    e-foldings of the inflation. The probability distribution of
    boundary conditions prefers  the tunneling boundary
      condition to the no-boundary boundary condition.
    Furthermore, for large values of a model parameter related
      to the inflaton mass and the cosmological constant, the
      probability of boundary conditions selects
      an unique boundary condition different from the tunneling
      type.
  \pacs{98.80.Qc, 98.80.Cq, 04.60.-m}
    \keywords{quantum cosmology; boundary condition; inflation;
      probability}
\end{abstract}

\maketitle
\section{Introduction}

Investigation of the very early period of the universe requires
quantum treatment of gravity~\cite{qua}. However, because we do not
have the complete theory of quantum gravity yet in hand,
simplified models with reduced dynamical degrees of freedom
have been investigated to understand nature of canonical quantum
gravity. This approach is the mini-superspace quantum cosmology (a
general review of quantum cosmology is given by
\cite{Halliwell:1990uy}).  A quantum state of the model is represented
by the wave function of the universe, which satisfies the
Wheeler-DeWitt (WD) equation derived from the procedure of canonical
quantization \cite{DeWitt:1967yk}. The wave function of the universe
is represented as the path integral by summing over histories of
the universe \cite{der}.

To obtain the wave function of the universe, we must impose
  boundary conditions of the WD equation.  In the context of the
quantum cosmology, there are two major candidates for the boundary
condition, the tunneling proposal by Vilenkin
\cite{cre,Vilenkin:1986cy} and the no-boundary boundary condition
proposal by Hartle and Hawking \cite{wave}. The former is given by the
wave function only consisting of the outgoing mode at the asymptotic
future of mini-superspace, and is analogous to the tunneling wave
function in quantum mechanics.  The latter is given by the path
integral  over Euclidean non-singular compact geometries with
no-boundary.
The path integral representation of the wave function provides
 important notions such as  analytic continuation of integration
contours, complex action and complex Euclidean solutions called the
fuzzy instantons \cite{Hartle:2007gi,cla}.  They play important rolls
when we consider semi-classical evaluation of the wave function of the
universe based on the saddle point method \cite{com}. A choice of the path
integral contour corresponds to specifying a boundary condition of the
WD equation \cite{ste,Halliwell:1989dy}.
To predict the classical universe using quantum cosmology, we
  must derive a probability for classical observables from the wave
  function of the universe. The number of e-foldings of inflation is
often used as a predictable observable and recent observational
restriction requires this number must be greater than about 60. The
amount of e-foldings predicted by quantum cosmology depends on
models and boundary conditions. Thus, the main issue in quantum
cosmology is to determine which type of boundary conditions are
preferable to explain observational results. Recent applications of
quantum cosmology to various cosmological models are studied in papers
\cite{Hwang:2013nja,Calcagni:2014xca,Sasaki:2013aka,Zhang:2014wia}.

  In this paper, we apply a numerical method to obtain
  predictions from wave functions of the universe. The main idea of
  our research is to represent boundary conditions of the wave
  function using exact solutions of the WD equation with a constant
  scalar field potential. This makes our problem of determining
  boundary conditions as the parameter estimation in space of boundary
  conditions.  We aim to obtain a probability distribution of
    boundary conditions under the constraint of sufficient e-foldings
    of the inflation.  This paper is organized as follows. In section
  II, we introduce a mini-superspace model and review derivaton of the
  probability for classical universes from the wave function of the
  universe.  In section III, we introduce a parametrization of
  boundary conditions and define the probability for boundary
  conditions.  Details of our numerical simulations and their results
  are explained in section IV.  Section V is devoted to summary and
  conclusion. We use the unit in which $c=\hbar=1$ throughout the
  paper.

\section{Mini-superspace model}

\subsection{Classical model and quantization}

We consider the Einstein gravity with a cosmological constant and
a minimally coupled massive scalar field as the inflaton. The action of the
gravity is given by
\begin{equation}
 S_G=\frac{1}{16\pi G}\int d^4 x\sqrt{-g}\,(R-2\Lambda),
\end{equation}
and the action of the scalar field $\Phi$ is
\begin{equation}
 S_m=-\frac{1}{2}\int d^4 x\sqrt{-g}\,[(\pa_\mu \Phi)^2 +m^2\Phi^2].
\end{equation}
We assume a homogeneous and isotropic closed universe.  Then the
geometry of the universe is given by the Friedmann-Robertson-Walker
(FRW) metric with a scale factor. We assume the following form of the metric:
\begin{equation}
 ds^2=\frac{3}{\Lambda}\left(-\frac{N^2}{q}d\lambda^2 + q\,d\Omega^2_3\right),
\end{equation}
where $\lambda$ is a dimensionless time parameter and $N$ is a lapse
function. We introduce a dimensionless field variable $\phi$ and its
mass $\mu$ as
\begin{equation}
  \phi=\left(\frac{4\pi G}{3}\right)^{1/2}\Phi,
\quad \mu=\left(\frac{3}{\Lambda}\right)^{1/2}m.
\end{equation}
Then the total action of our model becomes
\begin{equation}
 S=\frac{K}{2}\int d\lambda
 N\left[-\frac{1}{4}\left(\frac{q'}{N}\right)^2+q^2
   \left(\frac{\phi'}{N}\right)^2+1-q(1+\mu^2\phi^2)\right],
\end{equation}
where $'=d/d\lambda$ and we introduced a constant
$K \equiv 9\pi/(2G\Lambda)$.  In this model, dynamical variables are
the scale factor $q(\lambda)$ and the inflaton field
$\phi(\lambda)$. We represent them as coordinates of
configuration space
\begin{equation}
 q^A=(q^0,q^1)=(q,\phi).
\end{equation}
This configuration space is called mini-superspace.  The total
Hamiltonian of our mini-superspace model is given by
\begin{equation}
\label{eq:Ht}
H_T= \frac{KN}{2}\left[\frac{1}{K^2}\left(-4p_q^2 + \frac{1}{q^2}p_\phi^2
        \right)     -1 +q\left(1 + \mu^2\phi^2\right) \right]=NH.
\end{equation}
We obtain the Hamiltonian constraint by taking variation of the lapse
function $N$:
\begin{equation}
 H=0.
\end{equation}


Canonical quantization of the model is performed by replacing $p_A$ in
the Hamiltonian constraint by
differential operator $\hat{p}_A$ 
\begin{equation}
 p_A \rightarrow \hat{p}_A =-i\frac{\partial}{\partial q^A},
\end{equation}
and imposing operator version of the Hamiltonian constraint $\hat{H}$ on a
physical state $\Psi(q^A)$. It yields the Wheeler-DeWitt
equation
\begin{equation}
 \left[\frac{1}{2K^2}\left(4\frac{\partial^2 }{\partial q^2}
     - \frac{1}{q^2}\frac{\partial^2 }{\partial \phi^2 }\right)
            -\frac{1}{2} +q V(\phi)
 \right]\Psi(q,\phi)=0,\quad V(\phi)\equiv
 \frac{1}{2}+\frac{\mu^2}{2}\phi^2.\label{eq:wdwq0}
\end{equation}
In terms of $q^A$,
\begin{equation}
    \left[-\frac{1}{2K^2}G^{AB}\pa_A\pa_B+U(q)\right]\Psi(q)=0,\quad
    U=-\frac{1}{2}+qV,
    \label{eq:wdwq1}
\end{equation}
where $G^{AB}=\mathrm{diag}(-4,1/q^2)$ is a metric of
mini-superspace.  The wave function $\Psi(q^A)$ on mini-superspace is
called the wave function of the universe.  Although there is an
operator ordering ambiguity in the quantization procedure, 
 we choose the ordering which yields the equation
(\ref{eq:wdwq0}) in our analysis.

The wave function $\Psi$ can also be  expressed by the
path integral with respect to $q^A$ and $N$:
\begin{equation}
 \Psi(q^A) =\int \mathcal{D}N\mathcal{D}q^A
 \,e^{iS[N(\lambda), q^A(\lambda)]}, \quad
 S[N(\lambda),q^A(\lambda)]=\int_0^1d\lambda\,\mathcal{L}[N(t),q^A(t)].
\end{equation}
Here, $q^A\equiv q^A(1)$ is a boundary value of $q^A$ on the
final spacelike hypersurface $\lambda=1$. The path integral
representation of the wave function satisfies
Eq.~\eqref{eq:wdwq0}~\cite{der}. In our approach to the quantum
cosmology, we mainly focus on solving (\ref{eq:wdwq0})  as a
differential equation but the path integral representation plays an
important roll in characterizing boundary conditions for the wave
function.

\subsection{Semi-classical approximation and probability}
\subsubsection{WKB wave function}
To extract  predictions for the classical universe from the wave
function, it must be expressed as the semi-classical form, which means
the wave function behaves as the WKB solution of
Eq.~(\ref{eq:wdwq0}). We perform the WKB expansion of the wave
function as
\begin{equation}
 \Psi(q^A)=C(q^A)e^{-\frac{1}{\hbar}I(q^A)}.
 \label{eq:WKB}
\end{equation}
For convergence of the path integral representation of the wave
function, the contour of the path integral must be analytically
continued in the complex plane. Thus, $I$ and $C$ would
generally become complex functions. We write $I$ as $I=I_R-iS$ where
$I_R$ and $S$ are real functions.  We call $I_R(q^A)$ as a
pre-factor and $S(q^A)$ as a phase of the wave function.

Inserting \eqref{eq:WKB} into the WD equation
\eqref{eq:wdwq1}, we obtain a set of semi-classical equations for
$I, C$:
\begin{align}
O(\hbar^0) &:\quad -\frac{1}{2K^2}(\nabla I)^2 + U(q^A)=0  \label{eq:order1}, \\
O(\hbar^1) &:\quad 2\nabla I\cdot \nabla C +C\nabla^2 I=0 \label{eq:order2},
\end{align}
where $(\nabla I)^2=G^{AB}\pa_AI\pa_BI, ~(\nabla I)\cdot(\nabla
C)=G^{AB}\pa_AI\pa_BC, ~\nabla^2I=G^{AB}\pa_A\pa_BI$.
Now, we consider a condition of the wave function to provide
predictions for the classical universe. In the classical regime, the
phase of the wave function must satisfies the Hamilton-Jacobi
equation. This implies that the equation \eqref{eq:order1} 
corresponds to the  Hamilton-Jacobi equation in the classical
regime. Namely, 
\begin{equation}
 -\frac{ 1}{2}(\nabla I_R)^2 +i\nabla I_R \cdot \nabla S +\frac{
  1}{2}(\nabla S)^2 +K^2U =0
\end{equation}
should reduce to the classical Hamiltona-Jacobi equation
\begin{equation}
 \frac{1}{2K^2}(\nabla S)^2 + U =0.
\end{equation}
Thus $I_R$ and $S$ should satisfy the condition
\begin{equation}
\frac{ |\nabla I_R|^2}{ |\nabla S|^2} \ll 1 .   \label{eq:classic}
\end{equation}
This inequality is called the ``classicality''
condition~\cite{cla}. To predict the evolution of classical universes
from the wave function, $I,S$ and $C$ must satisfy
(\ref{eq:order1})-(\ref{eq:classic}) and we expect that the
probability can be obtained in the region of mini-superspace where the
classicality condition is satisfied.

In the path integral representation of the wave function, there are
more than one saddle point of the action in general. Thus, the semi-classical
wave function is given by superposition of WKB components
associated with  different saddle points:
\begin{equation}
 \Psi(q^A)=\sum_{i={\rm saddle}} C^{(i)}(q^A)e^{-I_R^{(i)}(q^A)} e^{iS^{(i)}(q^A)}.
 \label{eq:WKBwave}
\end{equation}
It is possible to obtain a desirable probability measure for the
classical universe using this
expression of the WKB wave function. 

\subsubsection{Conserved current and probability}
Now let us consider how to define probability from the wave
function. Introducing probability from the wave function of the
universe is not straightforward because the WD
  equation is the Klein-Gordon type and its conserved charge is not
  positive definite.  However, by considering the classicality
condition, it is possible to introduce a suitable probability measure
and we can define the conditional probability giving predictions for
observables.  For the wave function $\Psi$ satisfying the
WD equation, we have the following conserved current in
mini-superspace
\begin{equation}
 \mathcal{J}_A=\frac{i}{2}(\Psi^*\nabla_A\Psi-\Psi\nabla_A\Psi^*
 ),\quad
 \nabla \cdot\mathcal{J}_A  =0.
\end{equation}
In the classical region where the wave function has the WKB form, we
can obtain the positive definite probability measure from
$\mathcal{J}_A$. For each WKB components of the wave function
\eqref{eq:WKBwave}, we define
\begin{equation}
  J^{(i)}_A \equiv  -|C^{(i)}|^2 \exp(-2I^{(i)}_R) \nabla_A S^{(i)}.
\end{equation}
They are conserved independently in the classical region
$\nabla \cdot J^{(i)} =0$.  From the Hamilton-Jacobi equation, we can
assign the canonical momentum in the classical region
\begin{equation}
 p_A^{(i)} =\nabla_A S^{(i)}= \frac{ \partial S^{(i)}}{\partial q^A}.
 \label{eq:pHJ}
\end{equation}
Here, we focus only on the components with
$p^{(i)}_q=\partial_q S^{(i)}<0$. From
$dq^{(i)}/d\lambda\propto -p^{(i)}_q>0$, these components
correspond to expanding universes. Thus, we can introduce a conserved
current corresponding to expanding universes as
\begin{equation}
  J^+_A \equiv  -\sum_{p^{(i)}_q<0}|C^{(i)}|^2 \exp(-2I_R^{(i)}) \nabla_A S^{(i)} . 
\label{eq:exj}
\end{equation}
Let us consider a surface $\Sigma_c$ in mini-superspace which is
spacelike with respect to the metric $G_{AB}$ and has a unit
normal $n_A$. We require the classicality condition \eqref{eq:classic}
is satisfied on this surface. Then the relative probability
$\mathcal{P}(\Sigma_c)$ of classical histories passing through this
surface is given by the component of the conserved current
\eqref{eq:exj} along the normal if it is positive. In the
leading order in $\hbar$, this is
\begin{equation}
  \mathcal{P}(\Sigma_c) \equiv J^+ \cdot n = -\sum_{p^{(i)}_q<0}|C^{(i)}|^2
   \exp(-2I_R^{(i)})\nabla_n S^{(i)},
\end{equation}
where $\nabla_n$ means differentiation along the normal vector $n_A$.
As a point on $\Sigma_c$ is specified by the value of the scalar field,
$\mathcal{P}(\phi)\equiv\mathcal{P}(\Sigma_c(\phi))$ provides the
probability for the inflaton field to realize a value $\phi$ on $\Sigma_c$.

\subsubsection{Conditional probability for observables}

We can derive a probability for  observables from the probability
measure $\mathcal{P}(\phi)$. It can be given as the conditional probability
\cite{Halliwell:1990uy}
\begin{equation}
 P(s_0|s_1)=\frac{ \int_{s_0} J\cdot d\Sigma_c}{ \int_{s_1} J\cdot
  d\Sigma_c},\quad s_0\subset s_1,\label{eq:s2p}
\end{equation}
where $s_1$ is a subset of the hypersurface $\Sigma_c$ defined by some
theoretical constraints and $s_0$ is a subset of $s_1$ defined by
restricting $s_1$ using observational constraints.  By using the
relation \eqref{eq:pHJ}, we can obtain classical trajectories starting
from $\Sigma_c$. Namely, the probability measure on $\Sigma_c$ with
the classicality condition gives probability distribution of initial
data $(p_q,q, p_{\phi},\phi)$ for the classical equation of motion. In
our analysis, the number of e-foldings $\mathcal{N}$ is adopted as an
observable because this variable quantifies the inflationary models to
explain the horizon and the flatness problems. $\mathcal{N}$ is
defined by
\begin{equation}
 \mathcal{N} \equiv \log\left(\frac{a(t_{\text{f}})}{a(t_{\text{i}})}\right),
\end{equation}
where $t_{\text{i}}$ denotes the beginning time of inflation
and $t_{\text{f}}$ denotes the end time of inflation.  In our
analysis, we define $t_{\text{f}}$ as the end time of inflation
driven by the scalar field potential. The number of e-foldings
$\mathcal{N}$ is determined by the initial data and it is possible to
translate the probability measure for $\phi$ on $\Sigma_c$ to the
probability measure for $\mathcal{N}(\phi)$.

To introduce the conditional probability, we define an interval $s_1$
as $s_1=[\phi_\text{min},\phi_\text{pl}]$ where $\phi_\text{min}$ is
the lower bound of the interval and
$\phi_\text{pl}=4\sqrt{2K}/(3\mu)$ is the value of the
inflaton field corresponding to the Planck energy density
$m_\text{pl}^4$. Then an interval $s_0\subset s_1$ is defined as
$s_0=[\phi_{\rm suf},\phi_\text{pl}]$ where $\phi_{\rm suf}$
corresponds to the number of e-foldings
$\mathcal{N}_{\text{suf}}\approx 60$ consistent with observations.
Accordingly, the conditional probability to predict the universe with
sufficient inflation becomes
\begin{equation}
 P(s_0|s_1)=\frac{ \int_{\phi_{\text{suf}}}^{\phi_\text{pl}}d\phi\,
 \mathcal{P}(\phi)}{\int_{\phi_\text{min}}^{\phi_\text{pl}}d\phi\,
 \mathcal{P}(\phi)}.
\end{equation}
We denote this probability as
\begin{equation}
 P_{\rm suf}\equiv P(s_0|s_1)=P(\mathcal{N}\geq 60).
\end{equation}
The expectation value of $\mathcal{N}$  can be calculated as
\begin{equation}
 \langle \mathcal{N}\rangle =\frac{\int_{\phi_\text{min}}^{\phi_\text{pl}}d\phi
 \,\mathcal{N}(\phi)\mathcal{P}(\phi)}{\int_{\phi_\text{min}}^{\phi_\text{pl}}
 d\phi\, \mathcal{P}(\phi)}.
\end{equation}
These probability and expectation value depend not only on
cosmological models but also on boundary conditions of the wave
function. As we have already commented in the introduction, there are
two well known proposals for the boundary condition of the wave
function. One of them is ``no-boundary boundary condition proposal''
by Hartle and Hawking (HH), the other one is ``tunneling proposal'' by
Vilenkin (V). They predict different evolution of universe; (HH)
prefers small value of $\mathcal{N}$, on the other hands, (V) prefers
large value of $\mathcal{N}$. By calculating and comparing
$P_{\rm suf}$ for given models and given boundary conditions,
we can evaluate what type of models and boundary conditions are more
suitable to explain observation of our universe.

\section{Probability for boundary conditions}
When we have some restriction on our models of inflationary universe
from observations, we can investigate a probability which states
preferable type of boundary conditions. It is possible to express this
probability using  Bayes' theorem:
\begin{equation}
  P(B_i|S)=\frac{ P(B_i)P(S|B_i)}{\sum_{k}P(B_k)P(S|B_k)},
\end{equation}
where $P(B_i|S)$ is a probability for $B_i$ under $S$ happened. Here, 
$B_i$ is some candidate of a boundary condition of the wave
function labeled by index $i$, and $S$ means the universe with sufficient
inflation, namely, $\mathcal{N} \geq 60$.  Thus,
$P(B_i|S)$ denotes the probability for $B_i$ under the sufficiently
inflated universe. On the contrary, $P(S|B_i)$ in the right hand side
is the probability for sufficient inflation under the boundary
condition $B_i$ and is equivalent to $P_{\rm suf}$ defined in the
previous section
\begin{equation}
 P(S|B_i)=P_{\rm suf}(B_i)=P(\mathcal{N} \geq 60 \,|B_i).
\end{equation}
As we do not have any information on the prior probability $P(B_i)$,
we assume that it is uniformly distributed. To represent different
boundary conditions, we will introduce two parameters $a,b$ in
\eqref{eq:solWDW1}.  The probability for the parameters $a,b$ is given by
\begin{equation}
P(a,b|S)=\frac{P(S|a,b)}{\int da'db'P(S|a',b')}.
\label{eq:Bayes}
\end{equation}

\begin{figure}
 \centering
   \includegraphics[width=0.4\linewidth,clip]{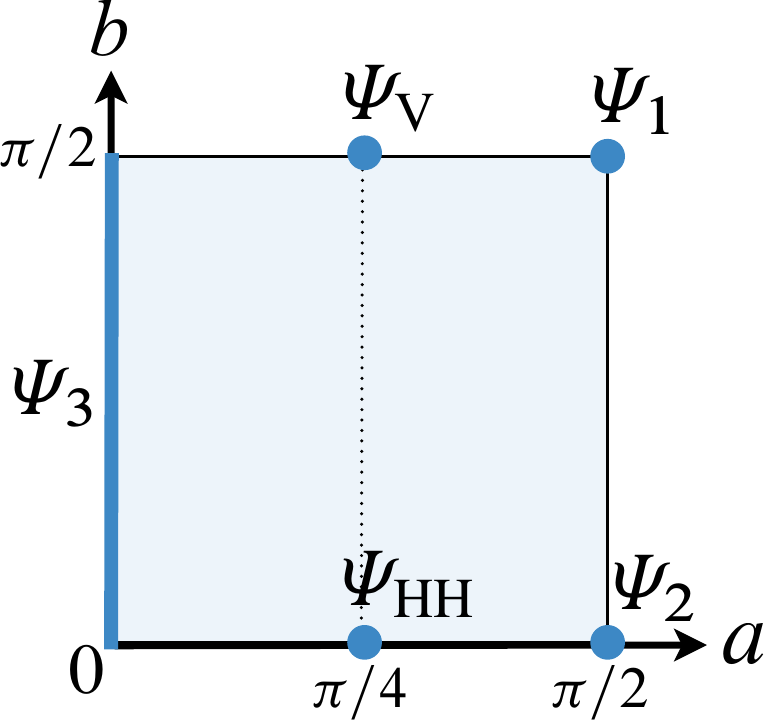}
  \caption{Parametrization $(a,b)$ of boundary conditions for $\Psi_C$.}
  \label{fig:paraBC}
\end{figure}

When we solve the WD equation, we have to impose some boundary
condition (in other words, initial condition) on the wave
function. For this purpose, we use exact solutions of the WD equation
which are obtained when the scalar field potential $V(\phi)$ is
constant.  Based on the path integral representation of the wave
function, for the constant scalar field potential case, the wave
function corresponding to the no-boundary (Hartle-Hawking) and the
tunneling (Vilenkin) type boundary conditions are expressed
as~\cite{ste}
\begin{equation}
  \Psi_{\text{HH}}=\Psi_2+\Psi_3,\quad 
  \Psi_{\text{V}}=\Psi_1+i\,\Psi_3,
\end{equation}
where
\begin{equation}
  \Psi_1\equiv (2V)^{-1/3}\mathrm{Ai}(z_0)\mathrm{Ai}(z),~\Psi_2
    \equiv (2V)^{-1/3}\mathrm{Bi}(z_0)\mathrm{Ai}(z),~\Psi_3\equiv
    (2V)^{-1/3}\mathrm{Ai}(z_0)\mathrm{Bi}(z),
\end{equation}
with
\begin{equation}
 z  =z(q)=\left(\frac{4V}{K}\right)^{-2/3}(1-2qV),\quad
 z_0=z(0)=\left(\frac{4V}{K}\right)^{-2/3}.
\end{equation}
For large values of the scale factor (classical region),
$\Psi_{\text{HH}}$ is superposition of expanding and contracting
universes with amplitude $\exp\left(+K/(6V)\right)$ which prefers
small values of the potential. On the other hand, $\Psi_{\text{V}}$
represents an expanding universe with amplitude
$\exp\left(-K/(6V)\right)$ which prefers large values of the
potential.  We can express more general type of  wave functions
introducing two real parameters $a,b$ which represent  boundary
conditions of the wave function
\begin{equation}
  \Psi_C = \tan a(\cos b\,\Psi_2-i\sin b\,\Psi_1)+\Psi_3,
  \quad 0\le a,b\le\pi/2.
\label{eq:solWDW1}
\end{equation}
Introduced parameters $a,b$ distinguish  boundary conditions of the
wave function (Table~\ref{ta:wave} and Fig.~\ref{fig:paraBC}).
\begin{table}
    \centering
    \caption{Typical wave functions and their parameters $(a,b)$ and
       asymptotic behaviors. The phase
      function $S_0$ is defined by
      $S_0=K/(6V)(2qV-1)^{3/2}-\pi/4$.}
    \begin{tabular}{|c|c|l|} \hline
      wave function & parameter $(a,b)$ &  asymptotic  form for
      $q\gg 1$\\ \hline
      $\Psi_{\text{HH}}$ & $(\pi/4,0)$
      & $ \sim\exp\left(+K/(6V)\right)\cos
                                        S_0$ \\ \hline
      $\Psi_{\text{V}}$ & $(\pi/4,\pi/2)$ & 
           $\sim\exp\left(-K/(6V)\right
                                                       )\exp(-iS_0)$
      \\ \hline
      $\Psi_1$ & $(\pi/2,\pi/2)$ &
     $\sim\exp\left(-K/(6V)\right)\cos
                                       S_0$ \\ \hline
      $\Psi_2$ & $(\pi/2,0)$ &
     $\sim\exp\left(+K/(6V)\right)\cos
                                       S_0$ \\ \hline
      $\Psi_3$ & $(0,\text{any values})$ &
     $\sim -\exp\left(-K/(6V)\right)\sin
                                       S_0$ \\ \hline
    \end{tabular}
    \label{ta:wave}
\end{table}
\noindent
Solving the wave function and calculating probability $P_{\rm suf}$
for different values of $(a,b)$, we can evaluate the probability for the
parameters $(a,b)$ using the relation \eqref{eq:Bayes}.

\section{Numerical simulation of the wave function}

\subsection{Boundary conditions and probability}
We solve the WD equation \eqref{eq:wdwq0} numerically to
obtain the probability of boundary conditions.  We prepare the initial
surface $q=q_\text{ini}$ in the Euclidean region of mini-superspace
and impose the following boundary condition for the wave function
$\Psi(q,\phi)$
\begin{equation}
  \Psi(q_\text{ini},\phi)=\Psi_C(q_\text{ini},\phi),\quad
  \pa_q\Psi(q_\text{ini},\phi)=\pa_q\Psi_C(q_\text{ini},\phi).
  \label{eq:WDBC}
\end{equation}
As $\Psi_C$ introduced by \eqref{eq:solWDW1} is specified by two
parameters $(a,b)$, this boundary condition is also specified by these
two parameters.  We call $\Psi_C$ the boundary wave function. For
models with a constant scalar field potential, this boundary
condition of course reproduces the exact solution $\Psi_C$.

\begin{figure}
 \centering
  \includegraphics[width=0.8\linewidth,clip]{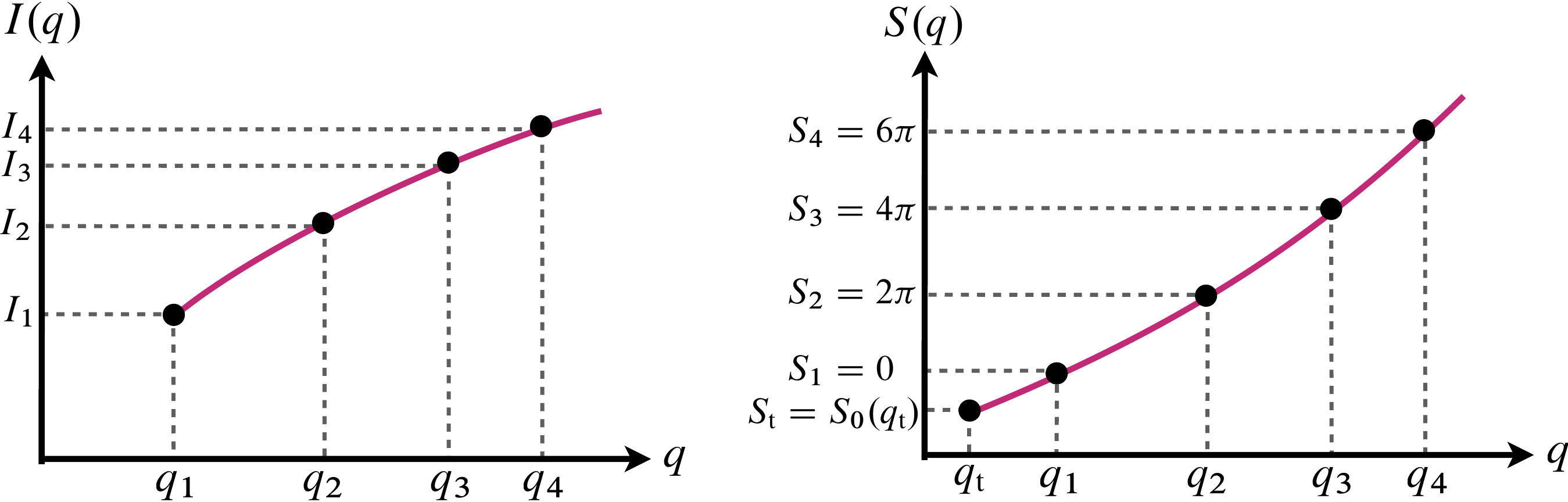}
  \caption{From the data set $[q_i, \Psi_i]$,  the prefactor $I_{R,I}$ and the phase function $S_{R,I}$ are determined by interpolation.}
  \label{fig:I_S_fit}
\end{figure}

In the Lorentzian region of mini-superspace with sufficiently large
value of $\phi$, the WD equation~ (\ref{eq:wdwq0}) has the
following asymptotic form
\begin{equation}
\left[ \frac{4}{K^2}\frac{ \partial^2}{\partial
	     q^2}-1+2qV(\phi)\right]\Psi(q,\phi)\approx0, \label{eq:qWDW}
\end{equation}
and the exact solution of this equation is given by
\begin{equation}
 \Psi_\infty = \alpha_1(\phi) {\rm Ai}(z)+\beta_1(\phi) {\rm
  Bi}(z),\quad z(q,\phi)=\left(\frac{4V(\phi)}{K}\right)^{-2/3}
(1-2q V(\phi)).
\end{equation}
Using the asymptotic form of the Airy function, $\Psi_\infty$ can be
expressed as superposition of two WKB modes corresponding to an
expanding universe and a contracting universe
\begin{equation}
 \Psi_\infty(q,\phi) \approx C_+(\phi) e^{-iS_0(q,\phi)}+C_-(\phi) e^{iS_0(q,\phi)},
\end{equation}
where $S_0$ is the phase function given by 
\begin{equation}
 S_0(q,\phi)=\frac{K}{6V(\phi)}\left(2V(\phi)q-1\right)^{3/2}-\frac{\pi}{4}.
\end{equation}
By fitting the numerically obtained wave function $\Psi_\text{num}$
with $\Psi_\infty$, we determine the prefactor of the WKB mode for the
wave function $\Psi_\text{num}$. Let us denote real and imaginary part
of the wave function $\Psi_\text{num}$ for a fixed value of $\phi$ as
\begin{equation}
  \left(\Psi_{\text{num}}(q)\right)_R= \exp[-I_R(q)]\cos S_R(q),\quad
  \left(\Psi_{\text{num}}(q)\right)_I=\exp[-I_I(q)]\cos S_I(q).
\end{equation}
From $(\Psi_{\text{num}})_{R,I}$, it is possible to determine
locations $q_i$ of local maximum points of the wave function and their
values $\Psi_i$. Thus, we obtain a set of data $[q_i, \Psi_i]$.
\begin{figure}
 \centering
  \includegraphics[width=0.4\linewidth,clip]{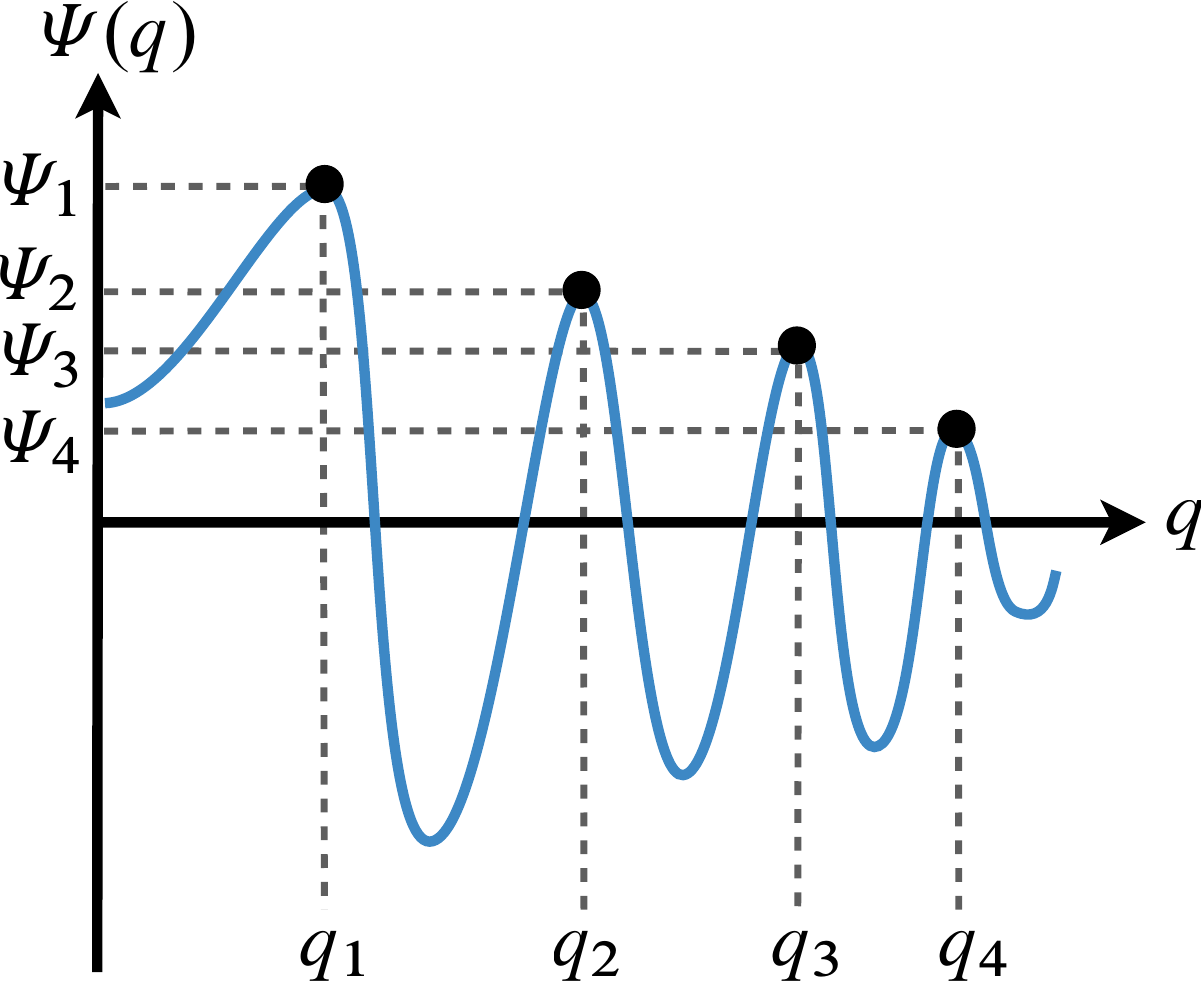}
  \caption{Local maximum points are determined from the numerical data of the
    wave function.}
  \label{fig:wave_fit}
\end{figure}
\noindent
Then, we obtain $(I(q))_{R,I}$ by interpolation of $(I_i)_{R,I}$ as
Fig.~\ref{fig:I_S_fit} (left panel).  We can also obtain phase
functions $S_{R,I}$ using a reference point $q_\text{t}=1/(2V(\phi))$
  with the asymptotic phase function $S_0$ as Fig.~\ref{fig:I_S_fit}
  (right panel).
\noindent
Repeating above procedures for other values of $\phi$, we can
determine a set of functions $(I_{R,I}(q,\phi), S_{R,I}(q,\phi))$
in the Lorentzian region.
 
Next, we introduce the phase difference relative to $S_0(q,\phi)$ from
the numerical data as
\begin{equation}
 \varphi_R=S_R-S_0, \hspace{5pt} \varphi_I=S_I-S_0.
\end{equation}
After that, we evaluate real and imaginary part of WKB amplitudes as
\begin{equation}
 C_{R+}=\frac{1}{2}e^{-I_R} e^{-i\varphi_R},\hspace{5pt} C_{R-}=\frac{1}{2}e^{-I_R} e^{i\varphi_R},\hspace{5pt} C_{I+}=i\frac{1}{2}e^{-I_I} e^{-i\varphi_I},\hspace{5pt} C_{I-}=i\frac{1}{2}e^{-I_I} e^{i\varphi_I}.
\end{equation}
Finally, we obtain amplitudes of the expanding and collapsing mode of
the WKB wave function as
\begin{equation}
 C_+=C_{R+}+C_{I+},\hspace{5pt} C_-=C_{R-}+C_{I-} \label{eq:pexpd}.
\end{equation}
The probability measure for the expanding universe  is
\begin{equation}
  \mathcal{P}(\phi)= -|C_+|^2\nabla_n S_0,
\end{equation}
where $n$ denotes a unit normal vector to a specified
spacelike hyperesurface $\Sigma_c$ in the classical region of mini-superspace.

To determine the hypersurface on which the probability is defined, we
must check the classicality condition \eqref{eq:classic} in our
simulation. The formal definition of the classicality is already
introduced in the section II, but applying it directly is not so easy
because decomposing the wave function to the phase function and the
prefactor is difficult.  However, as we also mentioned above, we
assume that the phase of the wave function can be well approximated by
the asymptotic phase function $S_0$ in the Lorentzian region. Thus, we
can define the desirable classicality condition for our simulation as
follows
\begin{equation}
  R_c \equiv \sum_{i=R.I} \frac{| (\nabla \tilde I_i)^2 +\nabla^2 \tilde I_i + i(2\nabla \tilde I_i \nabla S_0 -\nabla^2 S_0)|}{|(\nabla S_0)^2|} \ll 1, \label{eq:cl}
\end{equation}
where $\tilde I_i$ is defined by
\begin{equation}
   \tilde I_{R,I} \equiv I_{R,I} + i\,\varphi_{R,I}.
\end{equation}
When we calculate the probability for the classical universe, we
should choose a hypersurface $\Sigma_c$ with $R_c\ll 1$.


\subsection{Simulation set up}

\begin{figure}
  \centering
  \includegraphics[width=0.45\linewidth,clip]{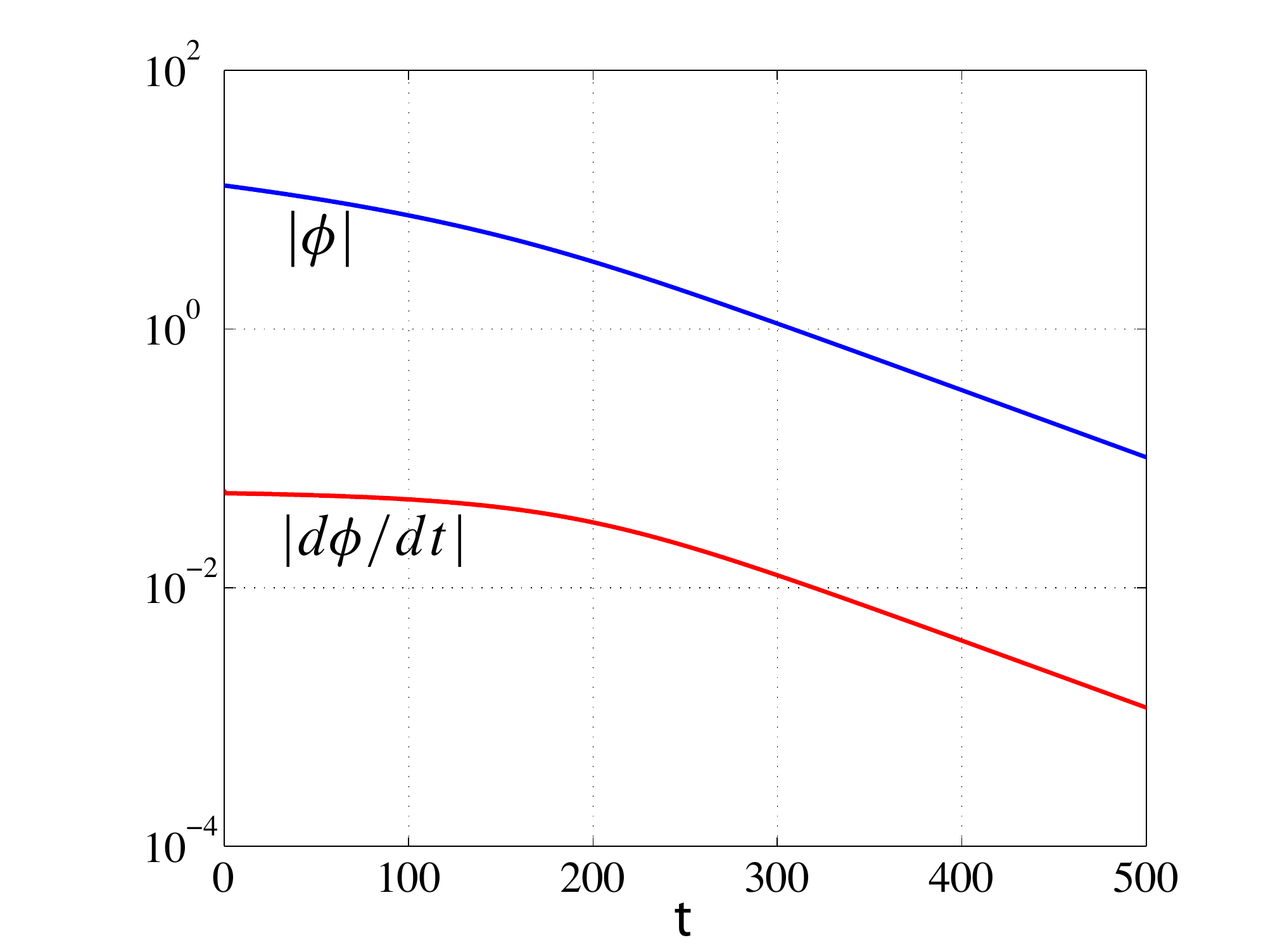}%
  \includegraphics[width=0.45\linewidth,clip]{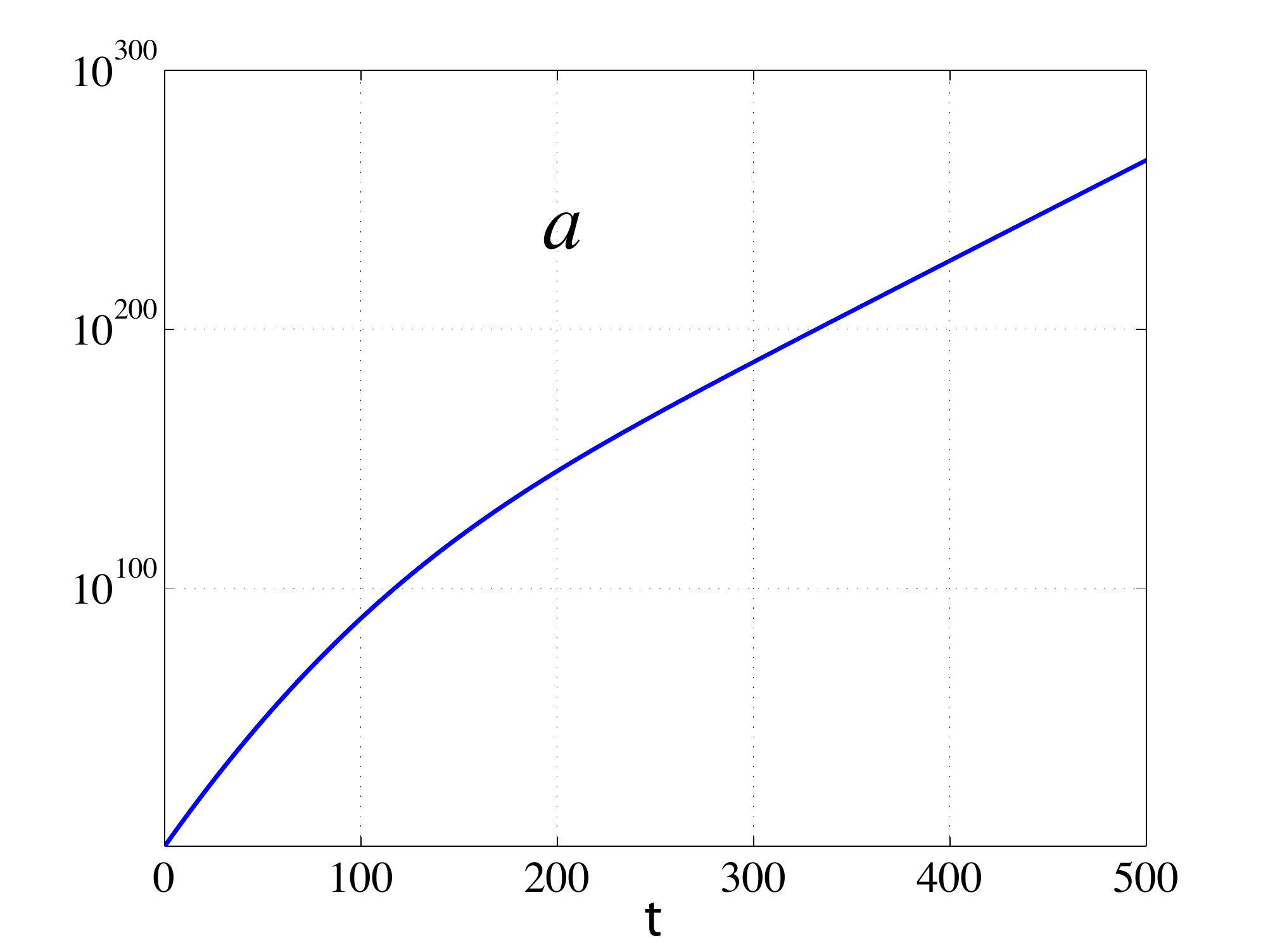}%
\vspace*{0.7cm} 
 \includegraphics[width=0.45\linewidth,clip]{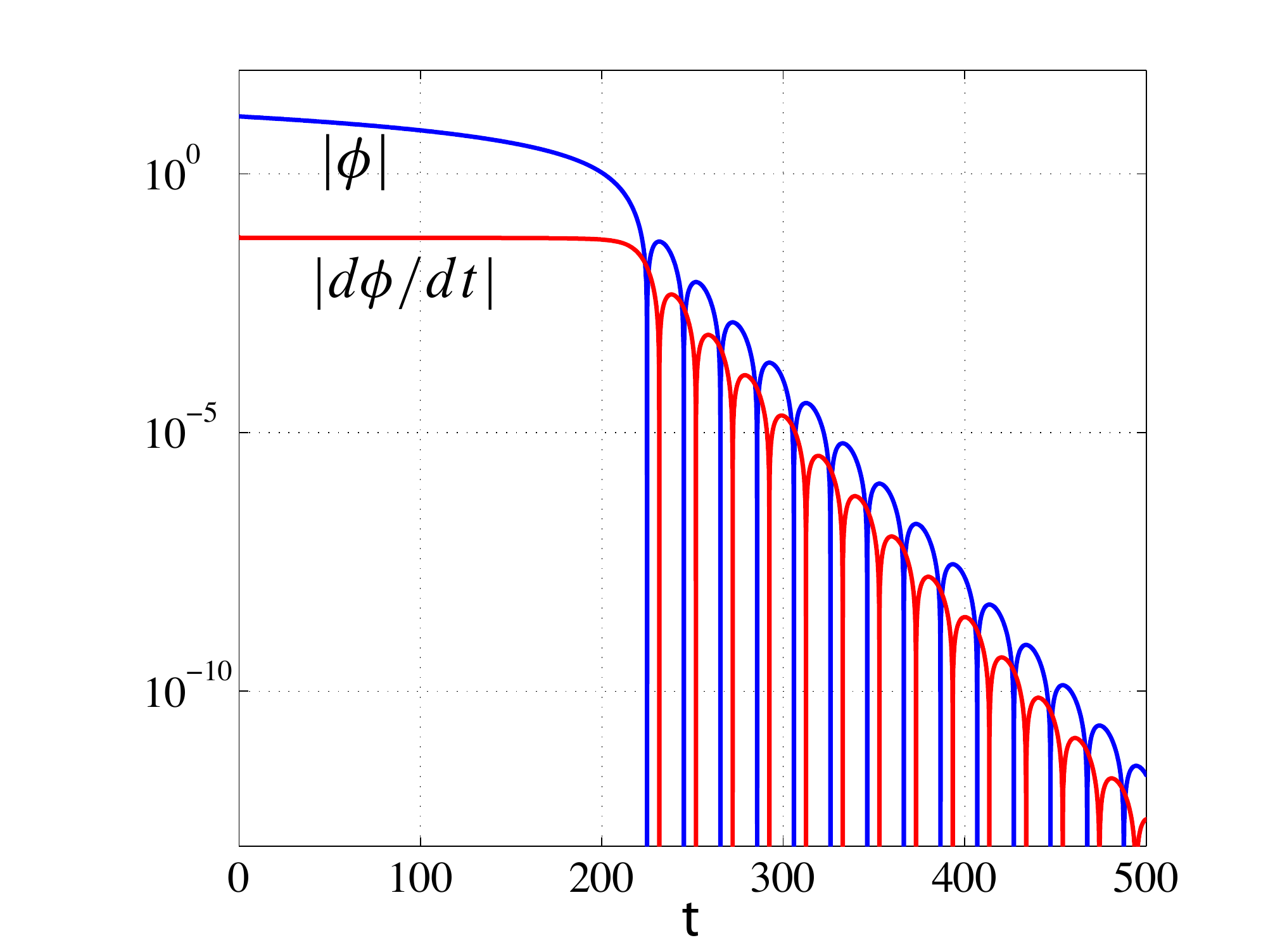}
  \includegraphics[width=0.45\linewidth,clip]{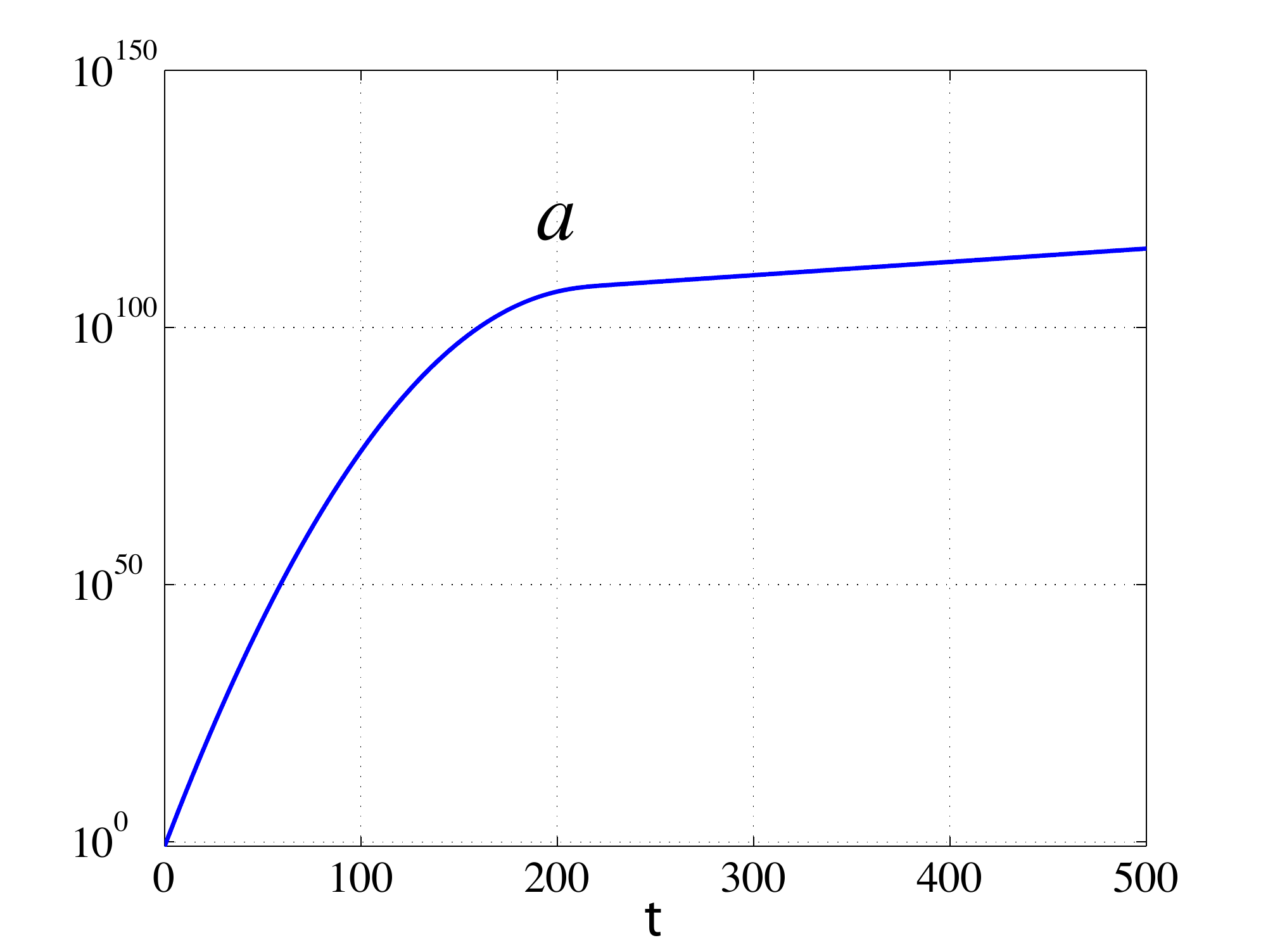}
  \caption{Classical evolutions of the inflaton $\phi(t)$ and the
    scale factor $a(t)$ for $\mu=0.2$ (upper panels) and $\mu=3$ (lower
    panels).}
    \label{fig:cltra}
\end{figure}

We fix the mass of the scalar field and consider two models with
parameters $\mu=0.2$ ($m^2=0.03$, $\Lambda = 2.25)$ and $\mu=3$
($m^2=0.03$, $\Lambda = 0.01)$. Different value of $\mu$
corresponds to different value of the cosmological constant in our
analysis. The former choice results in slow roll inflation followed by
over damped rolling of the inflaton field and the later results in
inflation with slow rolling followed by oscillation of the inflaton
about $\phi=0$. Samples of classical trajectory for these models are
shown in Fig.~\ref{fig:cltra}.
\noindent
The upper panels of Fig.~\ref{fig:cltra} shows a classical evolution
of $\mu$=0.2 model in terms of cosmic time $t$. The inflaton field
$\phi$ and its time derivative $\dot\phi$ decay monotonically and
inflation do not end (over damped oscillation). Until
$t_{\text{f}}\sim 200$ (in the unit of the Planck time), inflation is
driven by the mass term potential and after that time, inflation is
driven by the cosmological constant.  The lower panels of
Fig.~\ref{fig:cltra} shows a classical evolution of $\mu=3$
model.  In this model, $\phi$ decays and then oscillates with
exponentially damping at late time. The universe continues accelerated
expansion after the slow roll due to the cosmological constant.  Two
different behavior of classical solutions can be discriminated
by the dimensionless parameter $\mu$. For $\mu < \mu_*$,  classical
trajectories behave like the upper panels (over damped). For
$\mu > \mu_* $,  classical trajectories behave as the lower panels
(oscillation after slow roll).  The critical value  $\mu_*$ determined
by our simulation is $\mu_*\approx 1.5$.

Our simulation algorithm is as follows:
\begin{enumerate}
  \item Prepare an initial surface $q=q_\text{ini}$ in the Euclidean
  region of mini-superspace close to $q=0$. $q_\text{ini}$ cannot be
  chosen too small because we must keep the Courant condition for
  stable numerical integration of the wave equation. For the present
  case, the condition is
  \begin{equation}
    2q>\frac{\Delta q}{\Delta\phi},
  \end{equation}
 where $\Delta q$ and $\Delta \phi$ are grid spacings and
 $q_\text{ini}$ must satisfy this inequality.
 \item Solve the WD equation numerically from
 $q=q_\text{ini}$ to $q_{\text{fin}}$ with a given boundary wave
 function $\Psi_C$. We adopt the 5-step Adams-Bashforth method for
 numerical integration which has the 5-th order accuracy. We used
 $20000\times200$ grid size which covers
 $q_\text{ini}\le q\le q_\text{fin}, \phi_{\text{min}}\le\phi\le
 \phi_\text{max}$ (actual values used in the simulation is shown in
 Table \ref{ta:setup}).
\begin{table}[H]
    \centering
    \caption{Parameters of our simulation.}
    \begin{tabular}{|c|c|c|c|c|c|c|c|} \hline
      \diagbox{mass}{} & $K$ & $q_\text{ini}$ & ~$q_\text{fin}$~ & $\Delta q$ &  $~\phi_{\text{min}}~$  & ~$\phi_{\text{max}}$~ & $~\Delta\phi~$ \\ \hline
      $\mu=0.2$ & $6.283$ & $0.01$  & $14$ & $6.995\times 10^{-4}$  & $0$ & $26$ & $ 0.1307$ \\ \hline
      $\mu=3$ & $1413$ & $ 0.0001 $  &  $ 0.2 $  & $9.995\times 10^{-6}$ &  $1.8$ & $26$ &0.1216 \\ \hline
    \end{tabular}
      \label{ta:setup}
\end{table}
%
\item We specify a hypersurface $\Sigma_c$ on which the classicality
condition \eqref{eq:cl} is satisfied. We choose $\Sigma_c$ as a
constant $S_0$ surface. We numerically obtain the probability
$\mathcal{P}(\phi)$ on $\Sigma_c$.
 \item By integrating the classical equation of motion from $\Sigma_c$,
 we evaluate the number of e-foldings for  each classical
 trajectories. Then calculate the probability measure of the
 e-foldings.
\item Repeating step 2 to step 4 for different values of parameters
$(a,b)$, we obtain the probability of
parameters $(a,b)$ which specify  boundary conditions of the wave
function. We calculate the probability  for
$9\times 9$ grid points in the parameter space $(a,b)$ of  boundary
wave functions (Fig.~\ref{fig:gridBC}).
\begin{figure}
 \centering
   \includegraphics[width=0.4\linewidth,clip]{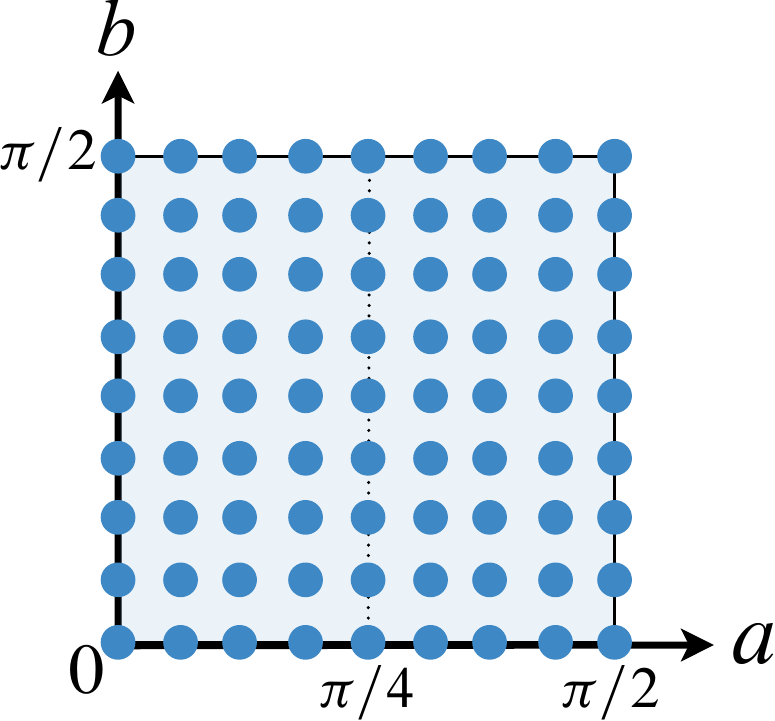}
  \caption{Parametrization of boundary conditions.}
  \label{fig:gridBC}
\end{figure}
\end{enumerate}

\subsection{Simulation results}

\begin{figure}
 \centering 
  \includegraphics[width=0.48\linewidth,clip]{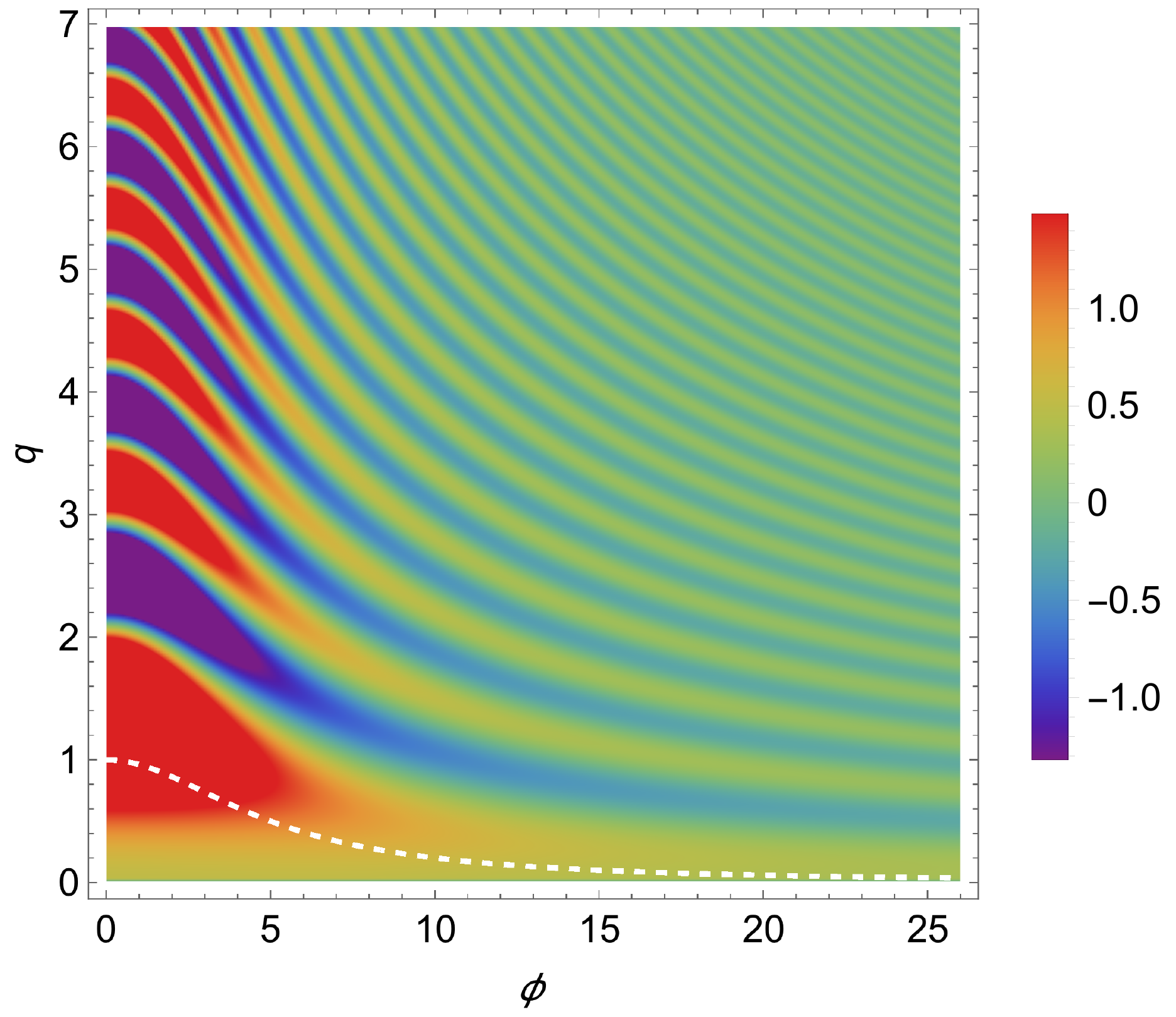}
 \includegraphics[width=0.5\linewidth,clip]{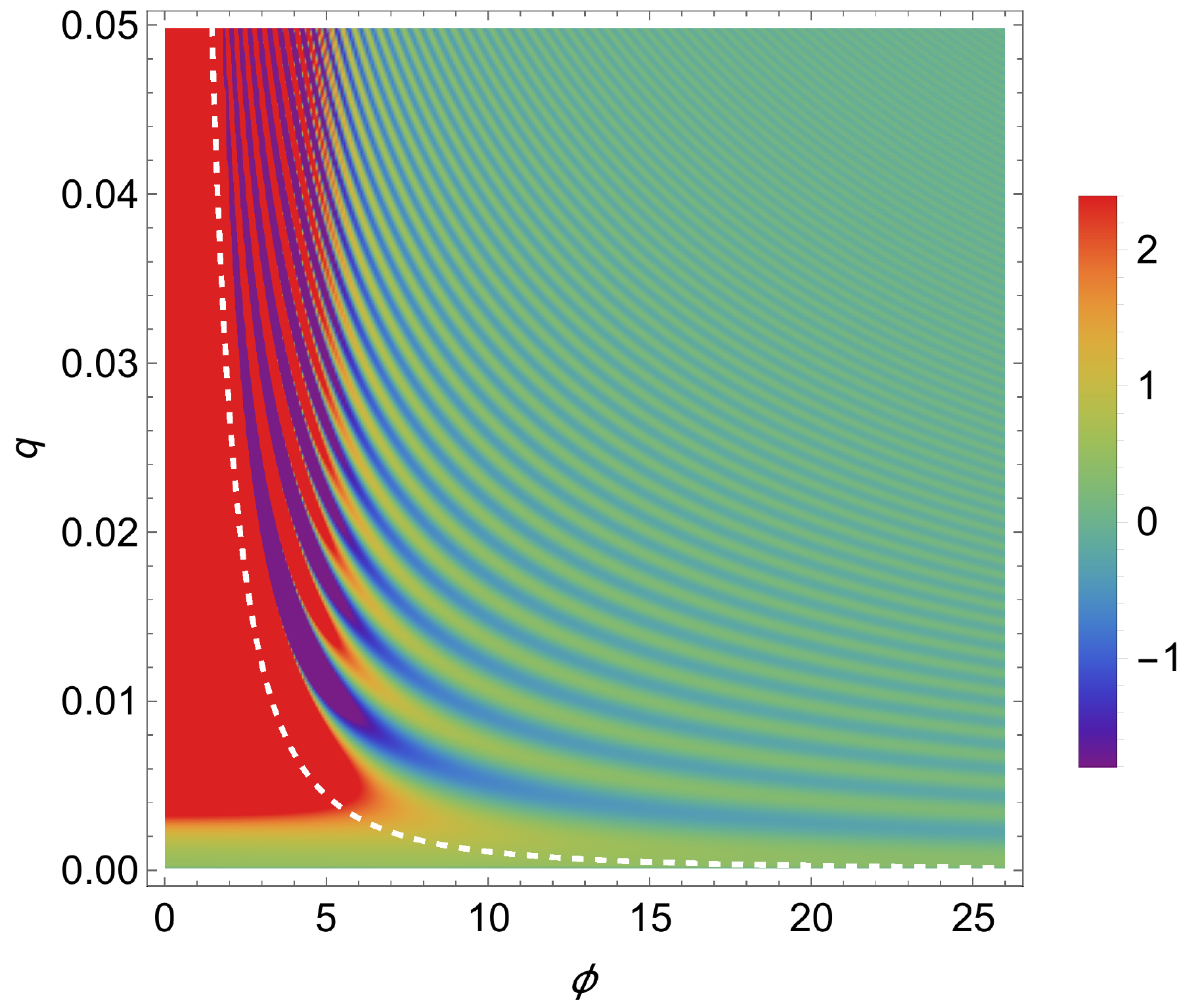}
 \caption{The density plot of  wave functions with the no-boundary
   boundary condition (HH). Left panel: $\mu=0.2$ ($m^2=0.03$,
   $\Lambda = 2.25),q_\text{ini}=0.01$. Right panel: $\mu=3$
   ($m^2=0.03$, $\Lambda = 0.01), q_\text{ini}=0.0001$. For this
   boundary condition,  wave functions are real. The dashed line
   represents $2 q V(\phi)=1$ which is the boundary between the
   Euclidean region and the Lorentzian region in mini-superspace.}
  \label{fig:wave_HH_R02}
\end{figure}

\begin{figure}
\centering
  \includegraphics[width=0.49\linewidth,clip]{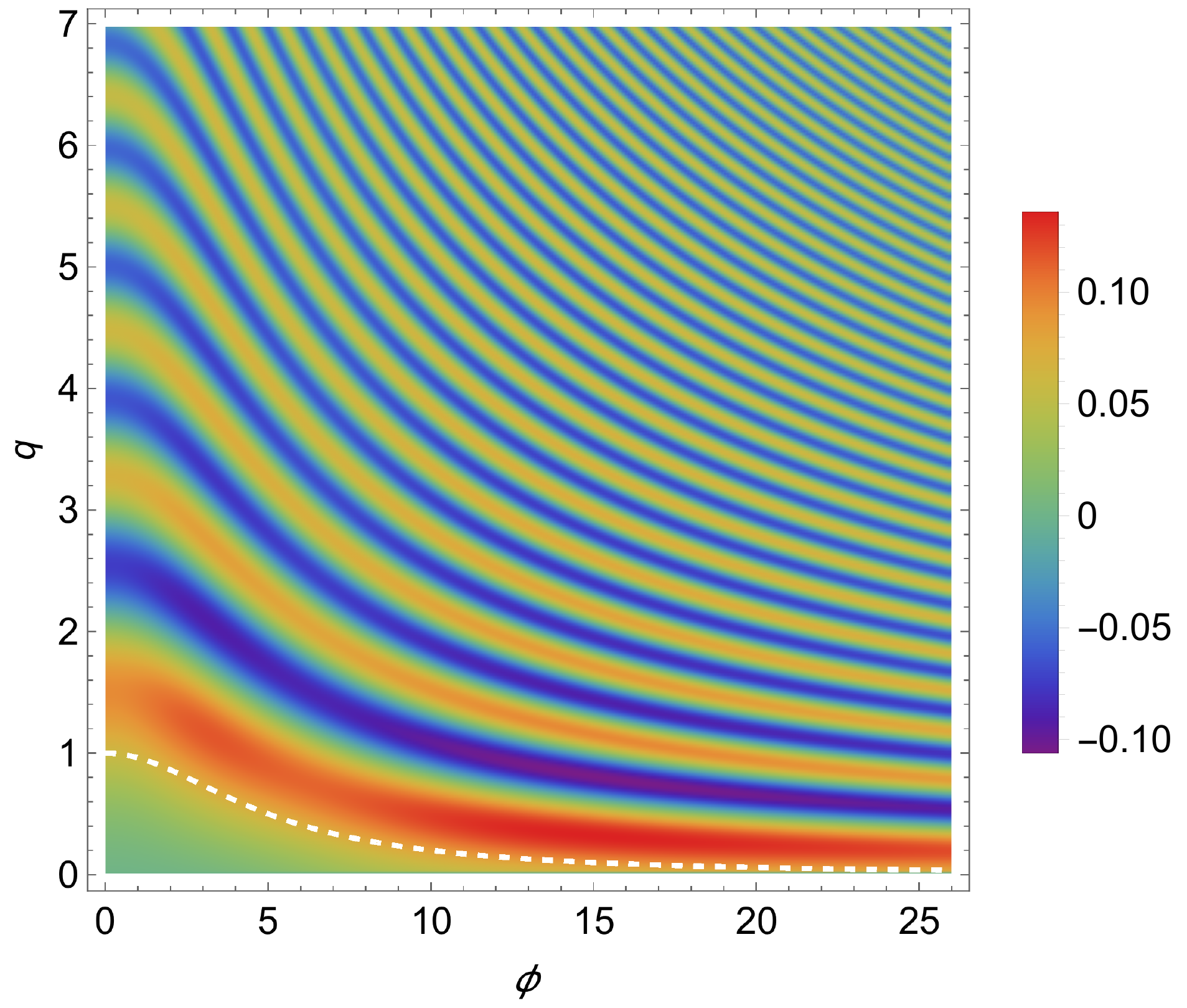}
  \includegraphics[width=0.49\linewidth,clip]{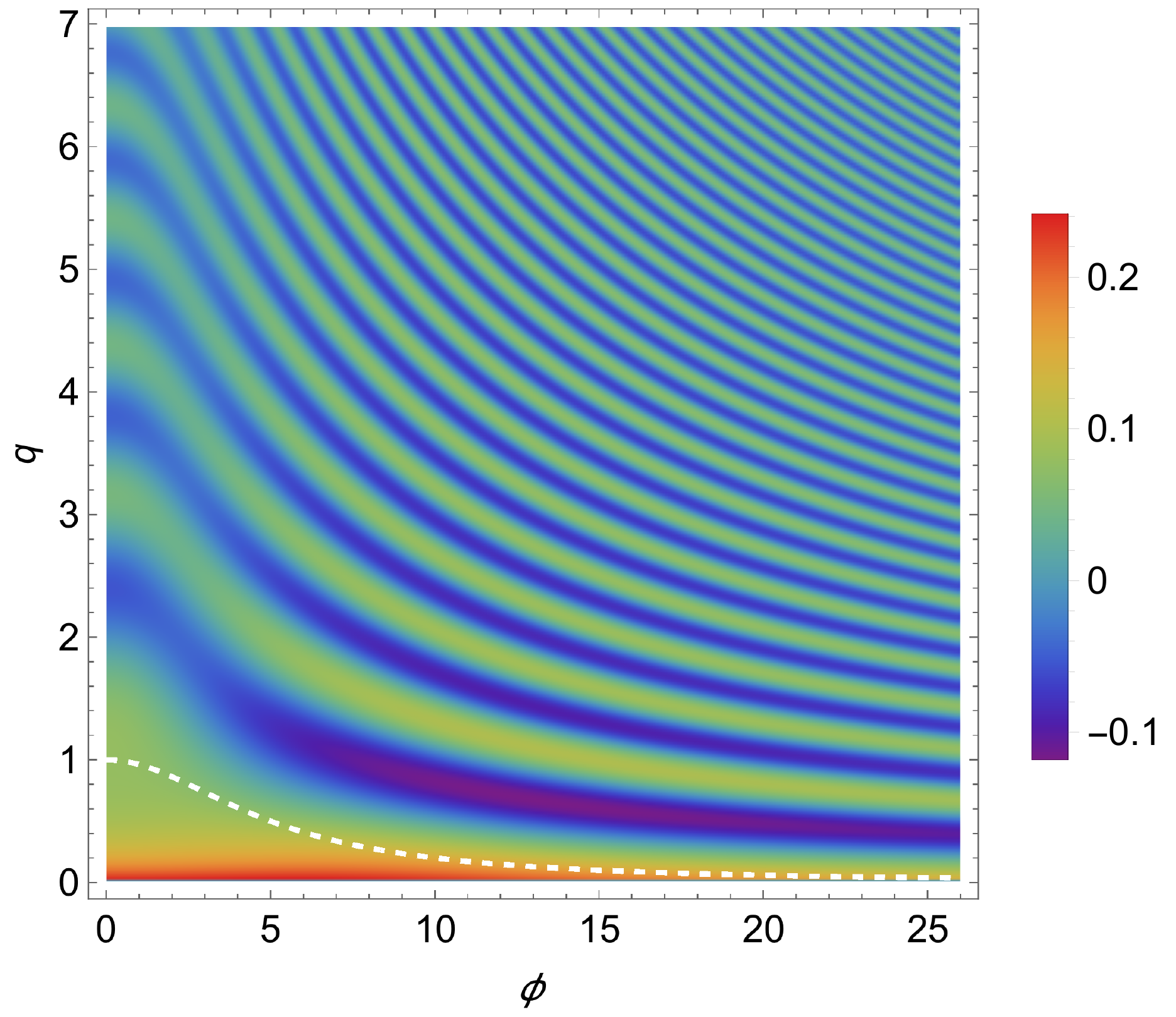}
     \caption{The density plot of the wave function with the tunneling
       boundary condition (V) for $\mu=0.2$, $q_\text{ini}=0.01$ (left:
       real part, right: imaginary part). The dashed line represents the
     boundary between the Euclidean region and the Lorentzian region.}
      \label{fig:wave_V_R02}
\end{figure}

\begin{figure}
  \centering
  \includegraphics[width=0.495\linewidth,clip]{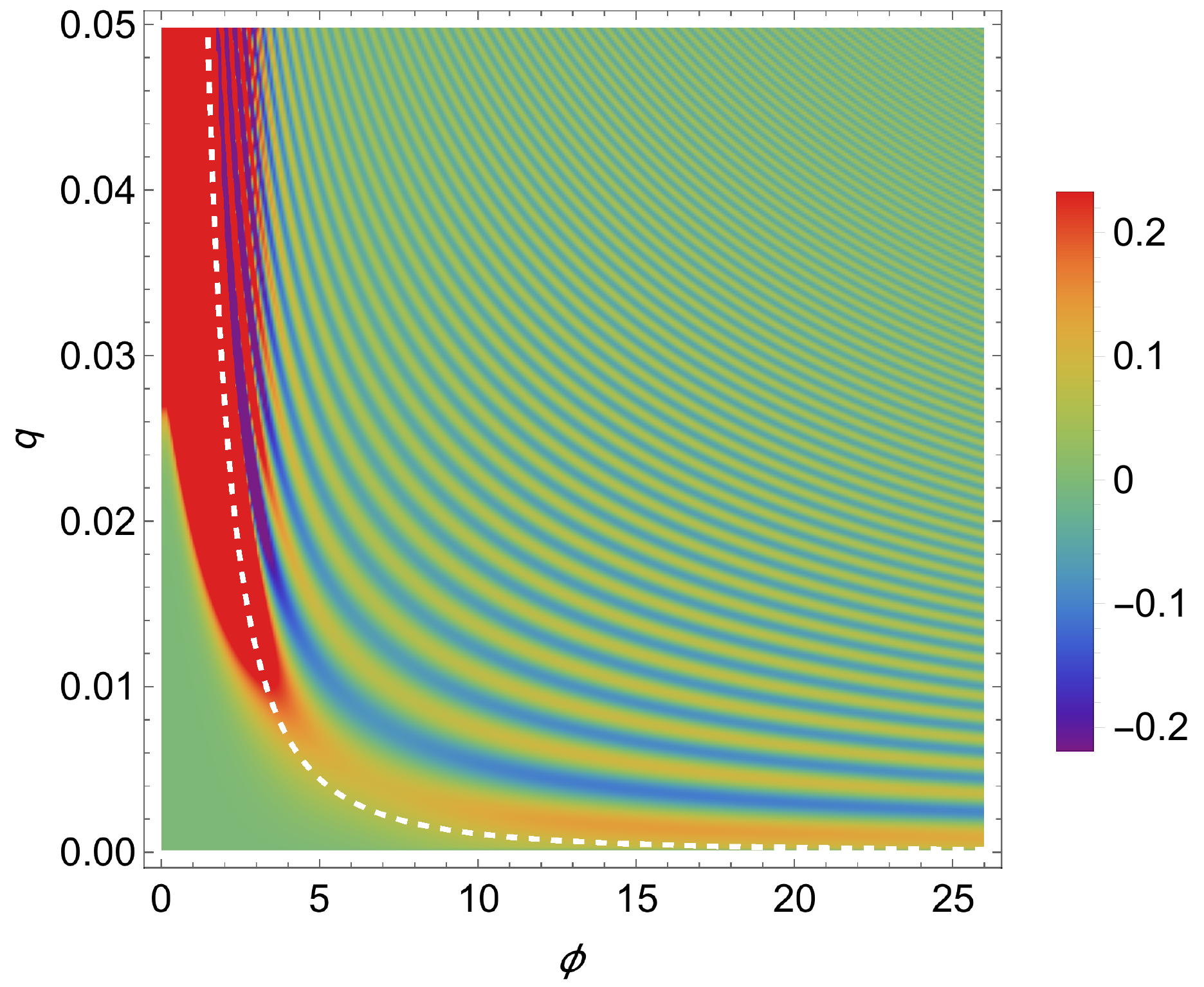}
  \includegraphics[width=0.495\linewidth,clip]{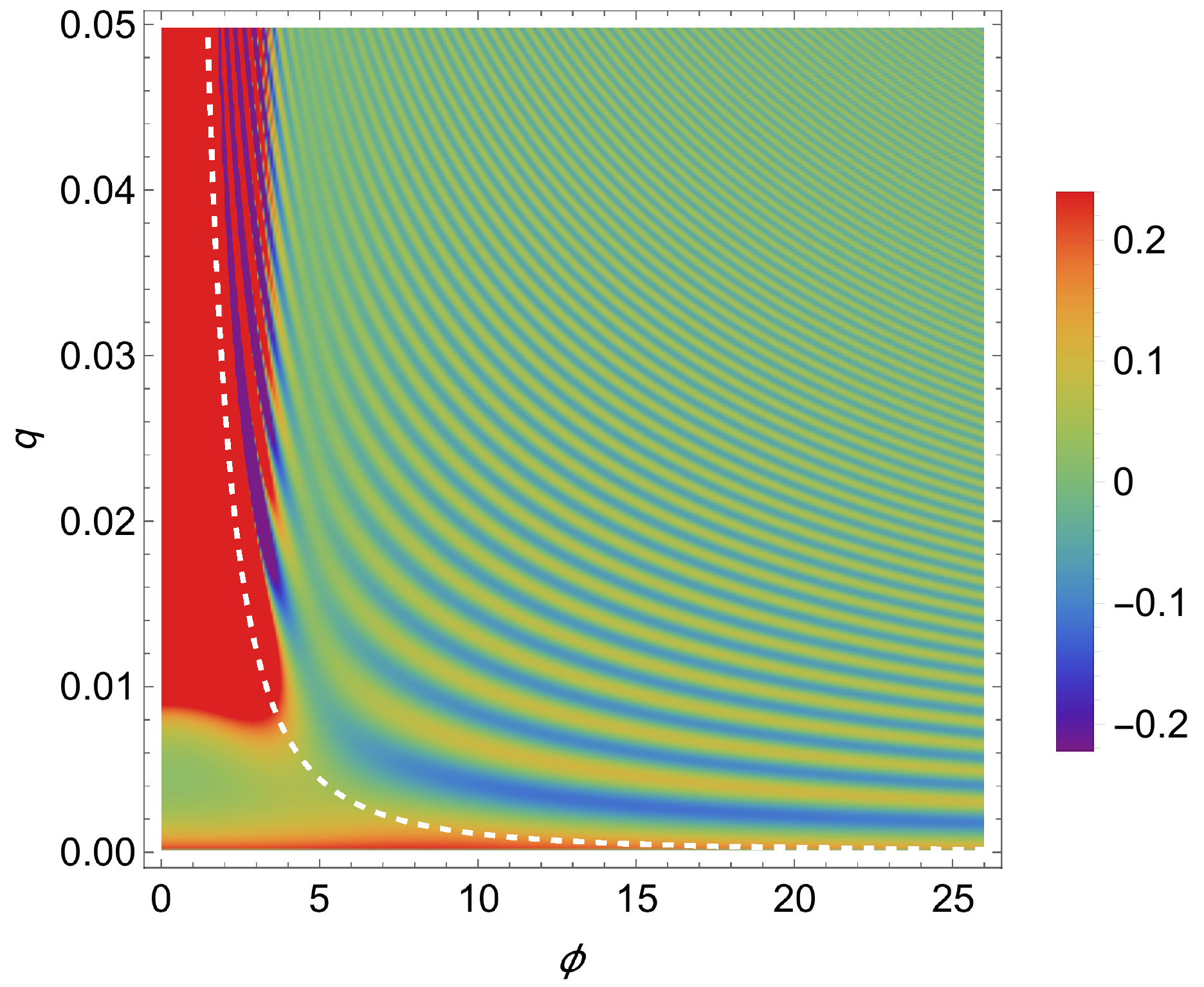}
  \caption{The density plot of the wave function with the tunneling
       boundary condition (V) for $\mu=3$, $q_\text{ini}=0.0001$ (left:
       real part, right: imaginary part). The dashed line represents the
     boundary between the Euclidean region and the Lorentzian region.}
      \label{fig:wave_V_R3}
\end{figure}

\noindent
Fig.~\ref{fig:wave_HH_R02}  shows
 wave functions with the boundary wave function $\Psi_\text{HH}$
(the no-boundary boundary condition (HH)). Fig.~\ref{fig:wave_V_R02}
and Fig.~\ref{fig:wave_V_R3} show  wave functions with the boundary
wave function $\Psi_\text{V}$ (the tunneling boundary condition (V)).
\begin{figure}
  \centering
  \includegraphics[width=0.47\linewidth,clip]{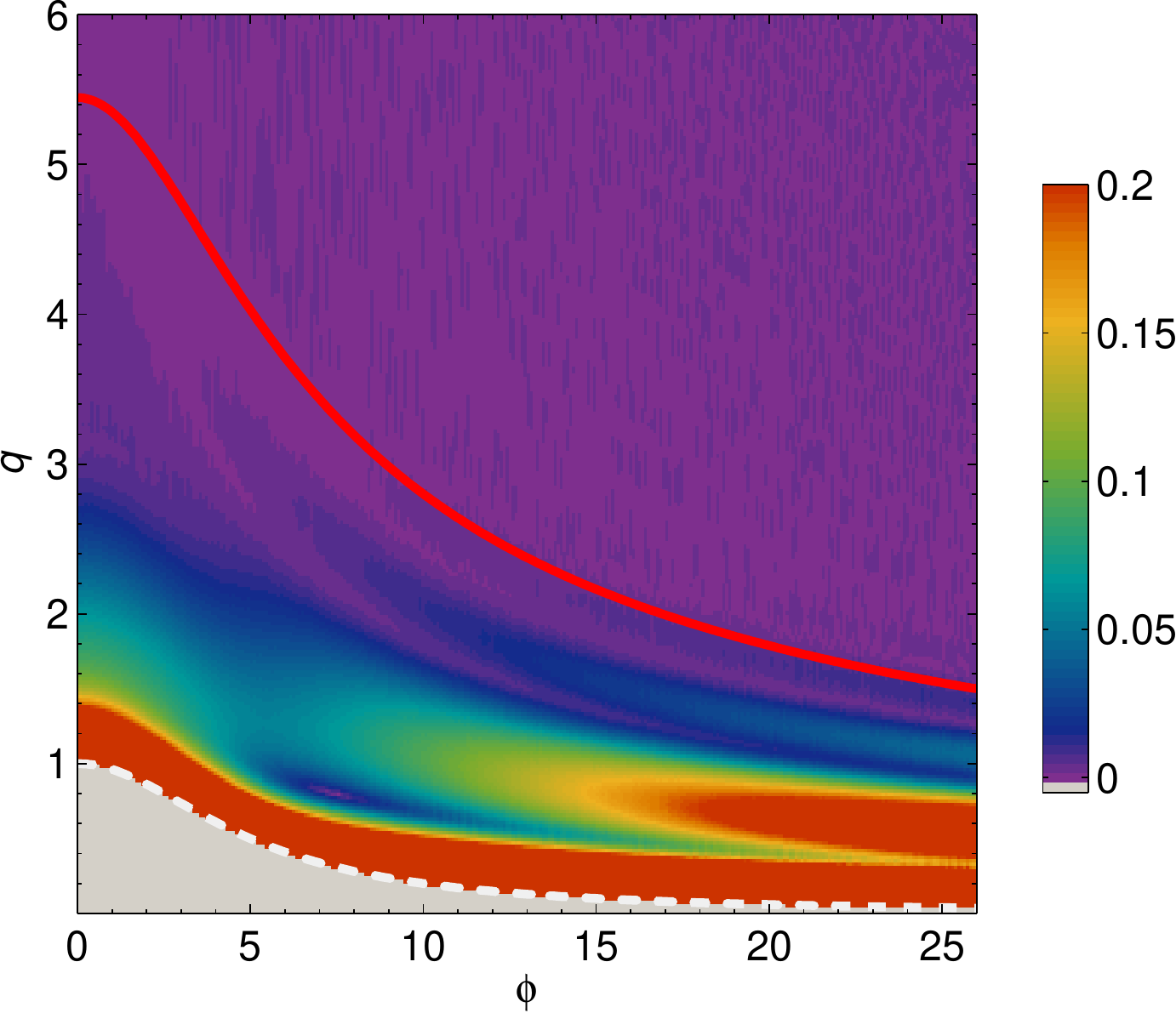}
 \includegraphics[width=0.49\linewidth,clip]{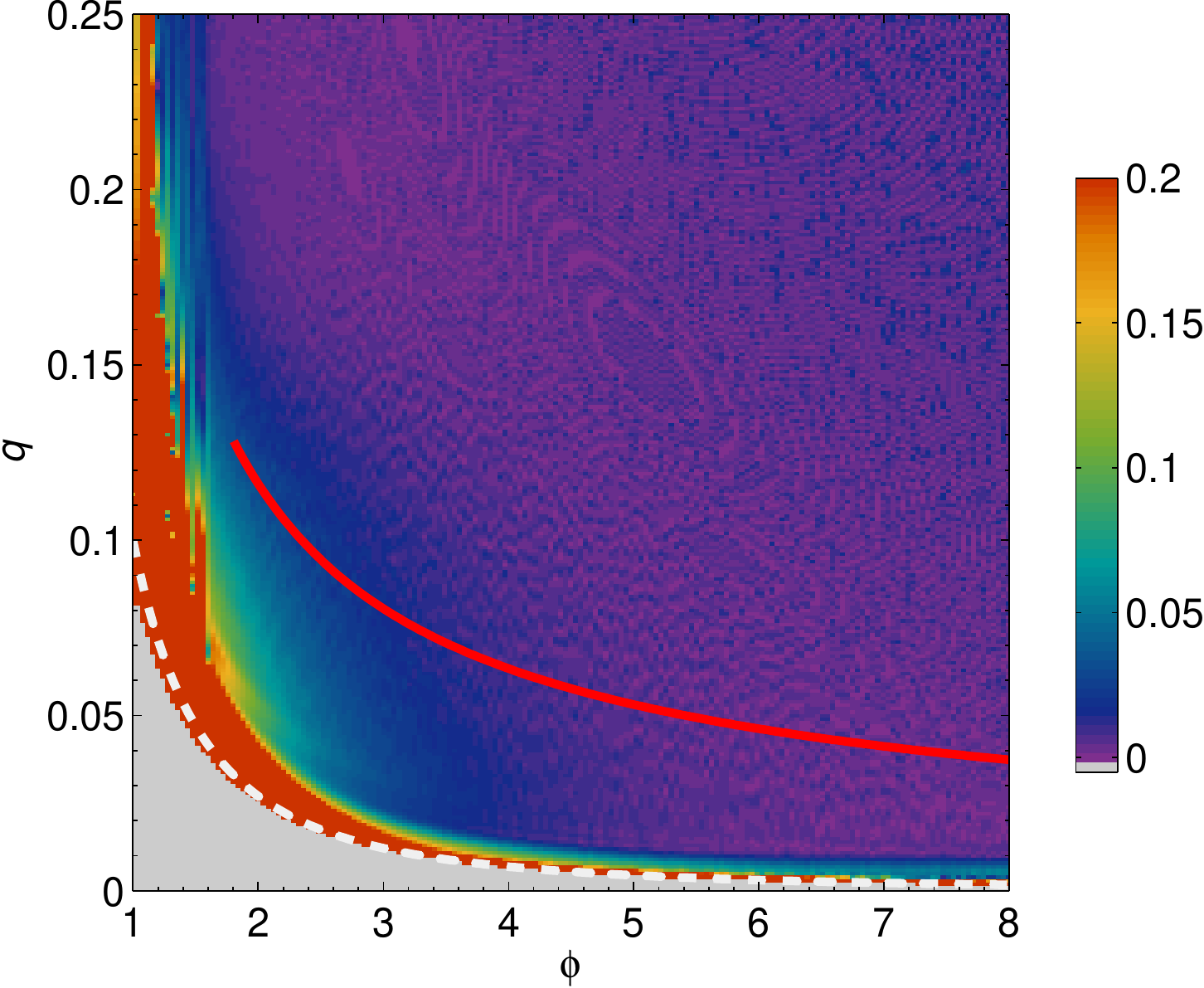} 
 \caption{The density plot of the classicality $R_c$ of the wave
   function with the no-boundary boundary condition (HH) (left:
   $\mu=0.2$, right: $\mu=3$). $\Sigma_c$ (solid line) is chosen as
   $S_0=\text{const.}$  in the region with $R_c<0.02$.  In the case of
   $\mu=3$, the classicality condition can not be satisfied for
   $\phi<1.8$. The dashed line represents the boundary between the
    Euclidean region and the Lorentzian region.}
       \label{fig:classic_1}
\end{figure}
Fig.~\ref{fig:classic_1} shows the classicality condition of the wave
function. We only show the case of the no-boundary boudnary condition
(HH) because the behavior of the classicality for other wave functions
is qualitatively same. We find out the region where the classicality
condition is satisfied and define the hypersurface $\Sigma_c$ in that
region.  In the case of $\mu=3$, the classicality condition can not be
satisfied for $\phi<1.8$ and we introduce a cut off
$\phi_{\text{min}}$ to calculate the conditional probability.  A
similar lower cut off of $\phi$ is already introduced
in~\cite{Hartle:2007gi,cla,Hwang:2013nja}.

\begin{figure}
  \centering
     \includegraphics[width=0.49\linewidth,clip]{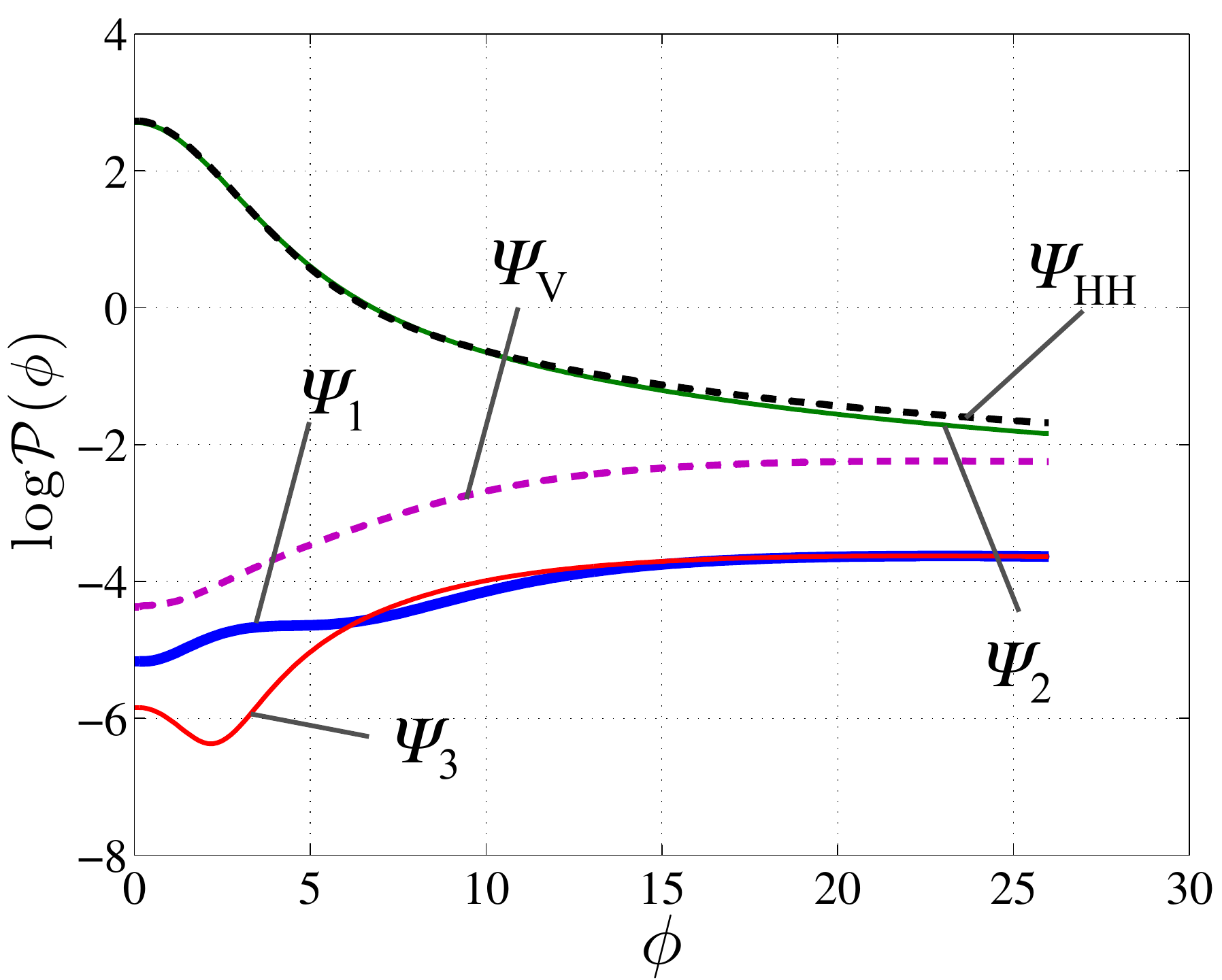}
     \includegraphics[width=0.49\linewidth,clip]{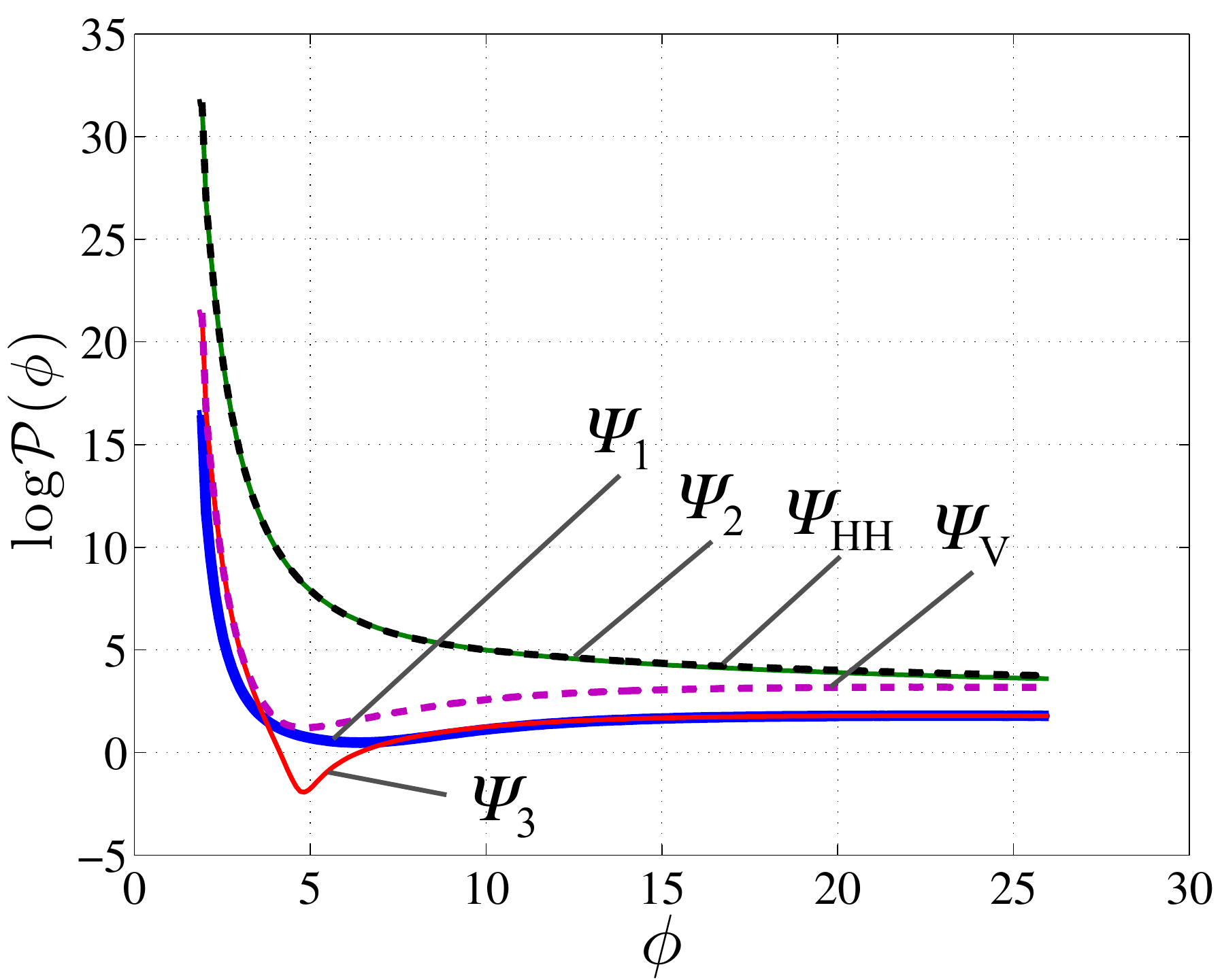}
     \caption{$\mathcal{P}(\phi)$ for $\mu=0.2$ (left) and
       $\mu=3$ (right). $\mathcal{P}(\phi)$ is not normalized.}
     \label{fig:P_phi_mu}
\end{figure}

Fig.~\ref{fig:P_phi_mu} shows $\mathcal{P}(\phi)$ obtained from
solutions of the WD equation obtained with boundary wave
functions $\Psi_1$, $\Psi_2$, $\Psi_3$ and $\Psi_{\text{HH}}$,
$\Psi_\text{V}$. From now on, we denote wave functions with
these boudary wave functions as
$\Psi_1,\Psi_2,\Psi_3,\Psi_\text{HH}, \Psi_\text{V}$. This probability
measure is not normalized because the conditional probability
$P(S|a,b)$ can be obtained without normalizing $\mathcal{P}(\phi)$.
\noindent
The left panel of Fig.~ \ref{fig:P_phi_mu} shows $\mathcal{P}(\phi)$
with each boundary conditions for $\mu=0.2$ model. Wave functions
$\Psi_1$, $\Psi_3$, $\Psi_{\text{V}} $ prefer large values of $\phi$
and $\Psi_2$, $\Psi_{\text{HH}}$ prefer small values of $\phi$. This
behavior of $\mathcal{P}(\phi)$ is the same as that obtained from the
exact wave function $\Psi_C$. However, the distribution for
small $\phi$ is slightly different from that obtained by
$\Psi_C$. A reason for this will be discussed soon later. The right
panel of Fig.~ \ref{fig:P_phi_mu} shows $\mathcal{P}(\phi)$ for
$\mu=3$ model. Probabilities for $\Psi_2$ and $\Psi_{\text{HH}}$ have
the same behavior as these in the  $\mu=0.2$ model. However,
probabilities for $\Psi_1$, $\Psi_3$ and $\Psi_{\text{V}}$ show
different behavior; The probabilities for small $\phi$ have
significantly large values for the $\mu=3$ model.
  
Here, we explain why behavior of $\mathcal{P}(\phi)$ for three wave
functions $\Psi_1$, $\Psi_3$ and $\Psi_{\text{V}}$ changes for $\mu=3$
in the small $\phi$ region. For $2qV(\phi)\ll1$, if we assume
$V(\phi)$ is constant, the wave function consists of two WKB modes
$\Psi_\pm \propto \exp(\mp (K/6V)(1-(1-2qV)^{3/2}))$.  In this region,
the wave functions $\Psi_3$ and $\mathrm{Im}(\Psi_\text{V})$ are
decreasing function of $q~(\propto\Psi_{-})$. On the other hand, the
wave functions $\Psi_1$ and $\mathrm{Re}(\Psi_\text{V})$ are
increasing function of $q~(\propto\Psi_{+})$ but their values are kept
small due to the prefactor $\exp(-K/(6V))$ (see Table \ref{ta:wave}).
The wave function $\Psi_3$ and $\mathrm{Im}(\Psi_\text{V})$ select the
decaying mode and their amplitudes are kept small until reaching the
boundary between the Euclidean and the Lorentzian region.  However,
$\phi$ dependence of the scalar field potential causes change of the
decaying mode $\Psi_-$ to the growing mode $\Psi_+$.  $\Psi_1$ and
$\mathrm{Re}(\Psi_\text{V})$ contain the growing mode with small
amplitudes and contribution of $\pa^2\Psi/\pa\phi^2$ term in the WD
equation becomes large around $q\sim q_0$ and enhances amplitudes of
their wave functions. As the result, amplitudes of $\Psi_1$ and
$\mathrm{Re}(\Psi_\text{V})$ acquire the similar distribution as
$\Psi_2$ and $\Psi_\text{HH}$ around the boundary between the Eulidean
and the Lorentzian regions. As $\Psi_\text{V}=\Psi_1+i\,\Psi_3$, all
wave functions have the similar distribution except their amplitudes.
This behavior of wave functions in the small $\phi$ region becomes
remarkable for large values of the model parameter $\mu$.  The mode
change and the growth of amplitude explained above may occur for any
values of $\mu$ (actually, occurs in both cases $\mu=0.2$ and
$\mu=3$).  If $\mu$ is small ($\Lambda$ is large), difference of
amplitudes between $\Psi_+$ and $\Psi_-$ is small and the enahancement
of amplitudes of $\Psi_1$ and $\mathrm{Re}(\Psi_\text{V})$ is not so
large. The mode change does not affect behavior of $\mathcal{P}(\phi)$
for small $\phi$.  But if $\mu$ is large ($\Lambda$ is small), the
difference and enhancement of the amplitude of the wave function
becomes remarkable and the behavior of $\mathcal{P}(\phi)$ for small
$\phi$ changes.

\begin{figure}
  \centering
     \includegraphics[width=0.49\linewidth,clip]{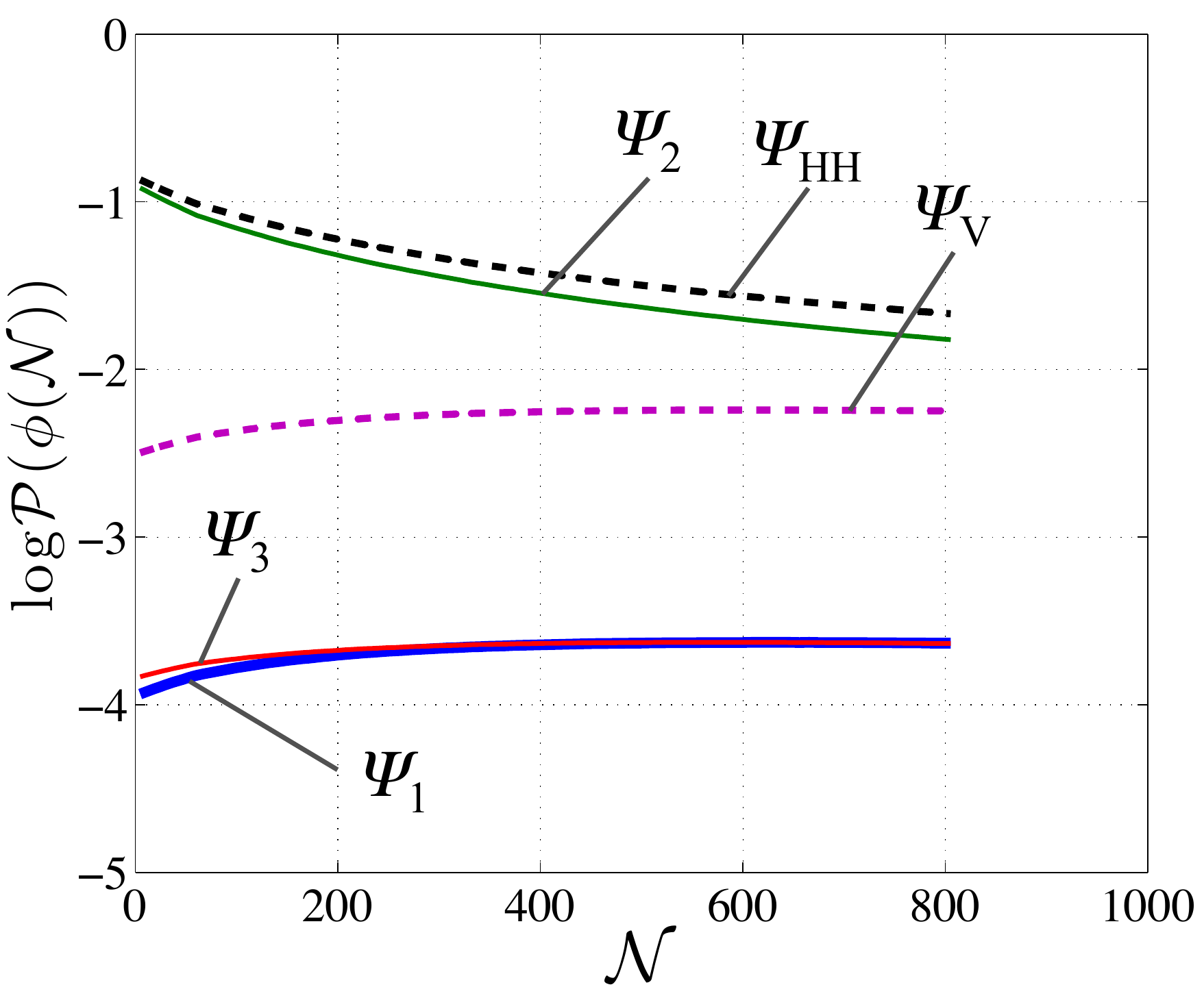}
     \includegraphics[width=0.49\linewidth,clip]{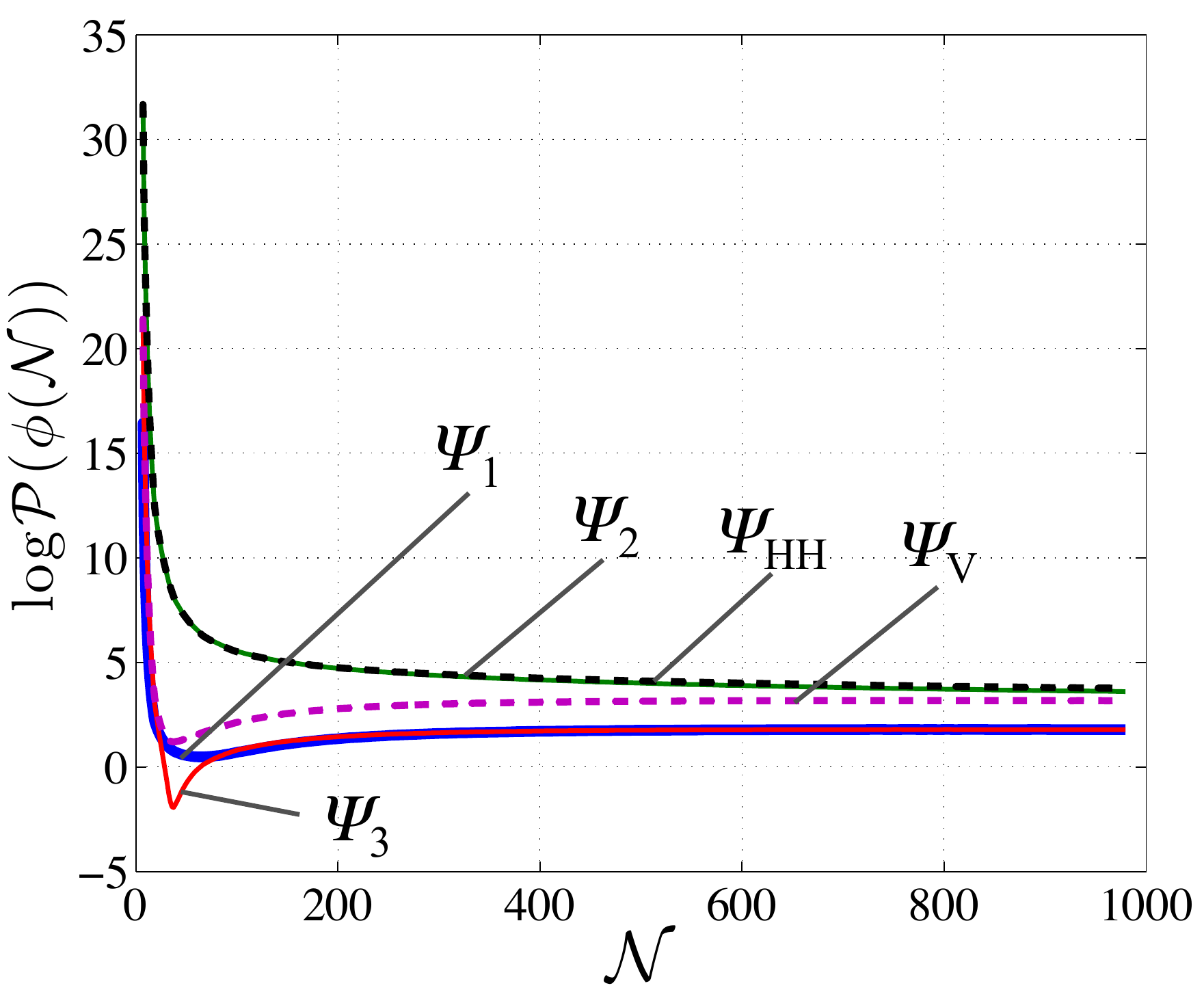}
     \caption{$\mathcal{P}(\phi(\mathcal{N}))$ for $\mu=0.2$ (left) and
       $\mu=3$ (right). $\mathcal{P}(\phi(\mathcal{N}))$ is not normalized.}
     \label{fig:P_N_mu}
\end{figure}

\begin{figure}
  \centering
     \includegraphics[width=0.45\linewidth,clip]{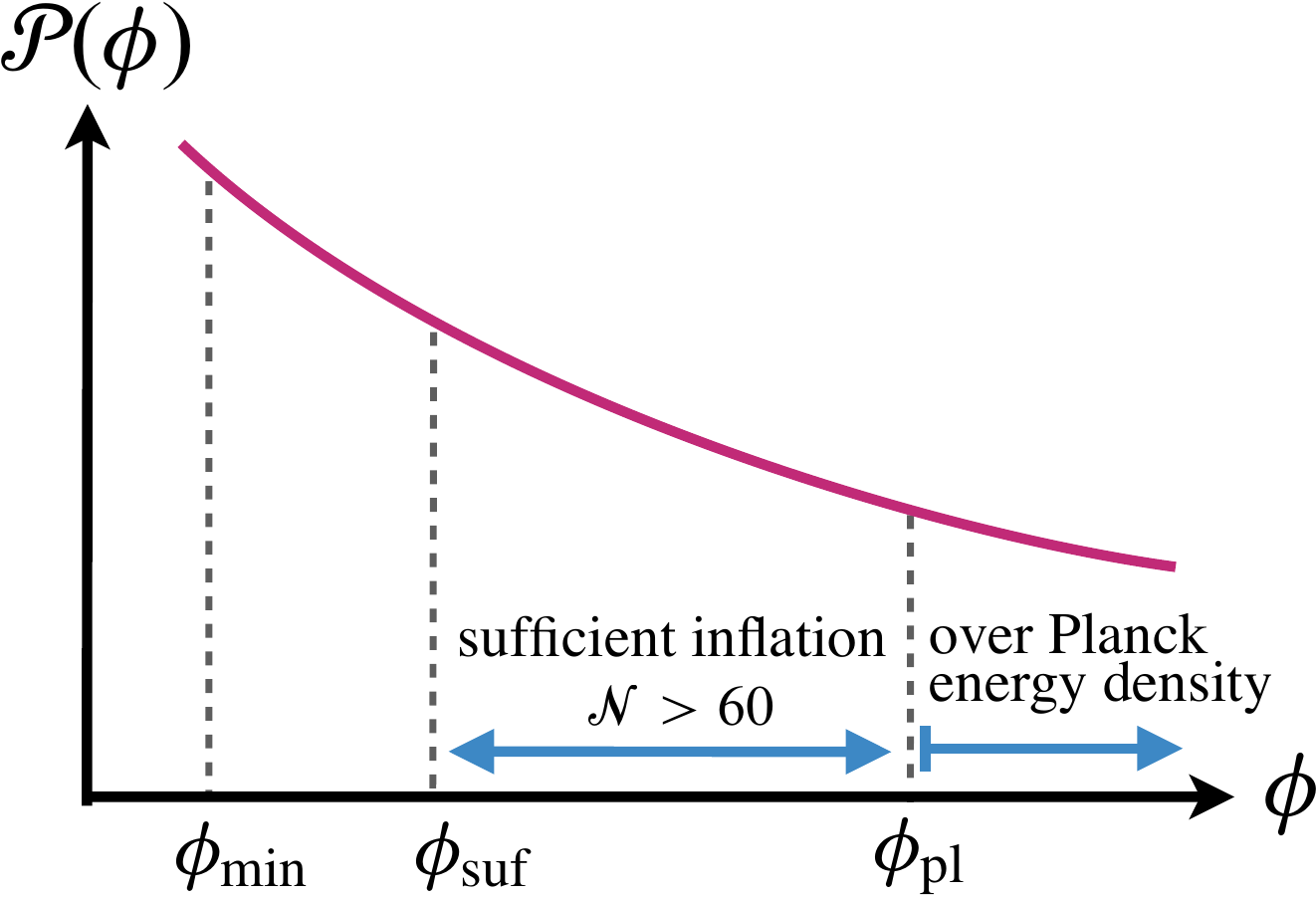}
     \caption{The probability is defined in
       $s_0 = [\phi_{\text{min}}, \phi_{\text{pl}}]$. We set
       $\phi_{\text{min}}=0$ for $\mu=0.2$ and $\phi_{\text{min}}=1.8$
       for $\mu=3$ because the classicality condition is violated in
       $\phi<1.8$ region for $\mu=3$ case.}
     \label{fig:prob_range}
\end{figure}

\begin{figure}
\centering
   \includegraphics[width=0.495\linewidth,clip]{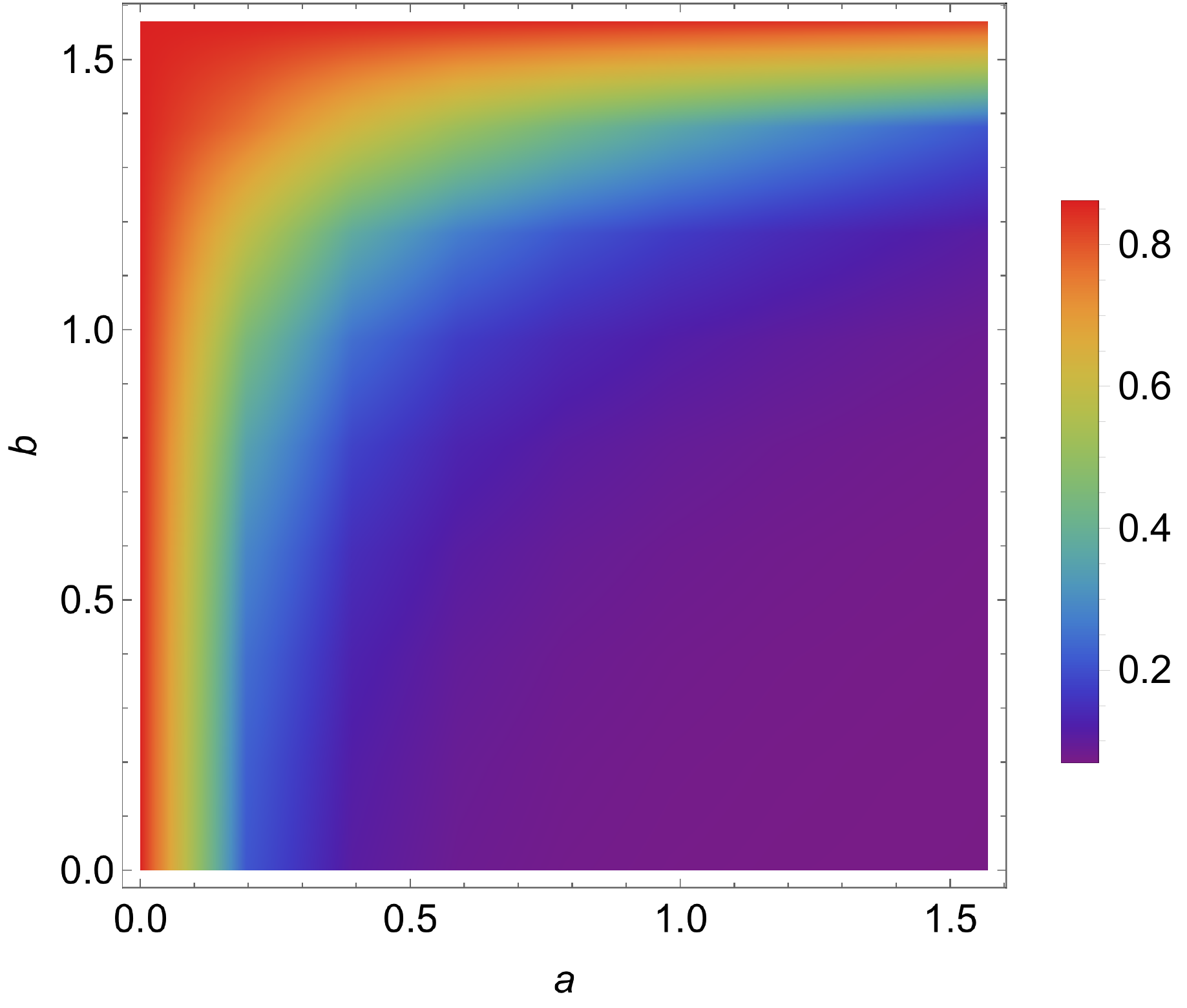}
   \includegraphics[width=0.495\linewidth,clip]{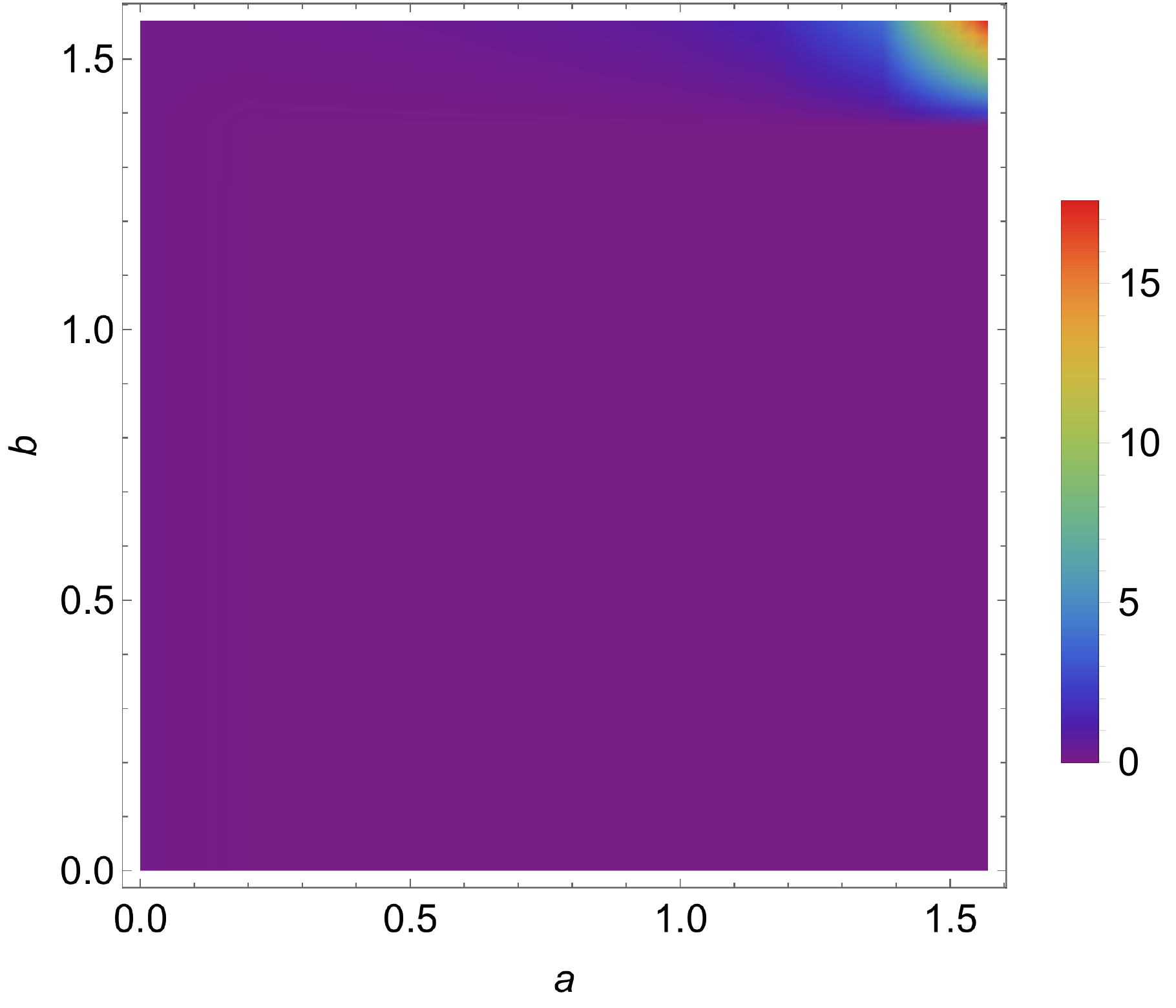}
   \caption{ Density
     plot of $P(a,b)$ for $\mu=0.2$ (left) and $\mu=3$ (right).
     We evaluated this probability on $9 \times 9$ grid points in the
     parameter space of boundary conditions.}
     \label{fig:P_ab_mu}
\end{figure}

Using $\mathcal{P}(\phi)$, we obtain the probability measure
for  number of e-foldings $\mathcal{N}$ by numerical integrations
of classical trajectories starting from $\Sigma_c$
(Fig.~\ref{fig:P_N_mu}).
\noindent
\noindent
Then, we evaluate the conditional probability for the sufficient
inflation $P_{\text{suf}} = P(\mathcal N\geq60)$. The numerical values
are shown in Table~\ref{ta:Psuf}, where we consider the probability in
the interval $s_0 = [\phi_{\text{min}}, \phi_{\text{pl}}]$
(Fig.~\ref{fig:prob_range}).
\begin{table}
    \centering
    \caption{$P_{\text{suf}}=P(\mathcal{N}\ge 60)$ for five wave
      functions
      and two choices of $\mu$. The larger value of $P_{\text{suf}}$
      is more  preferred for sufficient inflation. }
    \begin{tabular}{|c|c|c|c|c|c|} \hline
     {mass}$\backslash${wave function} & $\Psi_1$ & $\Psi_2$  & $\Psi_3$ &  
              $\Psi_{\text{HH}}$  & $\Psi_{\text{V}} $ \\ \hline
      $\mu=0.2$ & 0.604  & 0.0512 & 0.627 & 0.0561 & 0.621 \\ \hline
      $\mu=3$ & $2.40\!\times\!10^{-5} $ & $1.60\!\times\!10^{-10}$  &
                                                                       $1.72 
\!\times\! 10^{-7}$  & $1.61\!\times\!10^{-10}$ &
                                                  $6.74\!\times\!10^{-7}$
 \\ \hline
    \end{tabular}
      \label{ta:Psuf}
\end{table}
\begin{table}
    \centering
    \caption{ Expectation value of  number of e-foldings $\langle
      \mathcal{N} \rangle$ for five wave functions and two choices of
      $\mu$. This quantity represents amount of  inflation for the
      classical universe.}
    \begin{tabular}{|c|c|c|c|c|c|} \hline
      mass$\backslash$wave function & $\quad \Psi_1 \quad$ & $\quad\Psi_2\quad$  & $\quad\Psi_3\quad$ &  $\quad\Psi_{\text{HH}}\quad$  & $\quad\Psi_{\text{V}} \quad$ \\ \hline
      $\mu=0.2$ & 211  & 15.9 & 217 & 17.6 & 216 \\ \hline
      $\mu=3$ & $ 6.75 $  &  $ 6.74 $  & $6.74$ &  $6.74$ & $6.74$ \\ \hline
    \end{tabular}
      \label{ta:exN}
\end{table}
\noindent
Finally, we obtain the probability $P(a,b)$ of boundary conditions
using the relation \eqref{eq:Bayes} (Fig.~\ref{fig:P_ab_mu}).
\begin{table}
    \centering
    \caption{ $P(a,b)$ for five wave functions and two choices of
      $\mu$. }
    \begin{tabular}{|c|c|c|c|c|c|} \hline
      mass$\backslash$wave function & $\quad \Psi_1 \quad$ & $\quad\Psi_2\quad$  & $\quad\Psi_3\quad$ &  $\quad\Psi_{\text{HH}}\quad$  & $\quad\Psi_{\text{V}} \quad$ \\ \hline
      $\mu=0.2$ & 0.83  & 0.071 & 0.86 & 0.077 & 0.85 \\ \hline
      $\mu=3$ & 18  &  $1.2\times 10^{-4}$  & 0.13 & $1.2\times 10^{-4}$ &
                                                                       0.49
      \\ \hline
    \end{tabular}
      \label{ta:Pab}
\end{table}

In the case of $\mu=0.2$ (left panel of Fig.~\ref{fig:P_ab_mu}),
$P(a,b)$ has large values on lines $a=0$ and $b=\pi/2$. These two
lines correspond to the wave functions $\Psi_1$, $\Psi_3$ and
$\Psi_{\text{V}}$. $P(a,b)$ has small values on the line $b=0$,
corresponding to $\Psi_2$ and $\Psi_{\text{HH}}$. Consequently,
$\Psi_{\text{V}}$ is more preferable than $\Psi_{\text{HH}}$ to
realize large e-foldings, and this result is the same as
one predicted by the wave functions for a constant
scalar field potential. In contrast to that, in the case of $\mu=3$
(right panel of Fig.~\ref{fig:P_ab_mu}), $P(a,b)$ has large values
only around the point $(a,b)=(\pi/2, \pi/2)$, which corresponds to the
boundary wave function $\Psi_1$. This behavior is significantly
different from the case of $\mu=0.2$ (see
Table~\ref{ta:Pab}). Superiority of $\Psi_{\text{V}}$ to
$\Psi_{\text{HH}}$ holds both $\mu=0.2$ and $\mu=3$ models.  We expect
our results with parameter $\mu=0.2$ and $\mu=3$ are typical ones and
$P(a,b)$ with $\mu<\mu_*$ behaves similar to $\mu=0.2$ case and
$P(a,b)$ with $\mu>\mu_*$ behaves similar to $\mu=3$ case.

\section{Summary and conclusion}

 In this paper, we considered boundary conditions for the wave
  function of the universe which lead to sufficient e-foldings of
  inflation. For this purpose, we adopted the exact solutions of the
  WD equation with a constant scalar field potential as
  the boundary condition of the wave function, and solved the
  WD equation numerically. This boundary condition is
  parametrized with two real parameters and includes both the
  tunneling and the no-boundary boundary conditions. We obtained the
  probability distribution function for these parameters under the
  condition of sufficient e-foldings of inflation. The parameters with
  large value of this probability determines the boundary condition of
  the wave function which predicts sufficient e-foldings of
  inflation. We found that the probability distribution of boundary
  conditions has  two different behavior depending on the value of model
parameter $\mu$.

For small values of $\mu$, the cosmological constant dominates
  and the inflaton field asymptotically approaches to zero without
  oscillation. In this case, $\phi$ dependence of the wave function is
  not so strong and the obtained wave function reproduces behavior of
  exact wave functions with a constant scalar field potential. Hence
  the probability of boundary conditions has large values for
  $\Psi_1, \Psi_3, \Psi_\text{V}$ and small values for
  $\Psi_2, \Psi_\text{HH}$. Thus, boundary conditions
  $\Psi_1, \Psi_3, \Psi_\text{V}$ are preferable to realize sufficient
  period of inflation and superiority among them is small. This
  behavior of the probability of boundary conditios can be expected
  from behavior of wave functions for a constant scalar field potential. 
On the other hand, for large values of $\mu$, the slow roll
  inflation is followed by oscillation of the inflaton field. In this
  case, the derivative term of $\phi$ in the
  WD equation cannot be neglected and  wave functions
   have large values about $\phi=0$ for any boundary conditions. Owing
   to this behavior of the wave function, the probability of boundary
   conditions has large value about $\Psi_1$ (superiority of
   $\Psi_\text{V}$ over $\Psi_\text{HH}$ is kept as before). Thus, 
   realistic inflationary models followed by oscillation of inflaton
   field select the boundary condition $\Psi_1$.

As an extention of analysis presented in this paper, it is also
  possible to discuss probability for values of the model parameter
  $\mu$. The probability distribution of boundary conditions has a sharp
  peak for the model with large $\mu$ and selects a specific boundary
  condition. This implies that a suitable boundary condition is
  automatically chosen for large values of $\mu$ (small values of the
  cosmological constant). If we assume that the probability of
  boundary conditions select an unique boundary condition, the
  parameter $\mu$ must acquire large value (the cosmological
  constant must be small). To confirm this expectation, we should
  analyse behaviour of the probability of boundary conditions for
  wider range of the parameter $\mu$ and we will report on this
  subject in a separate publication.

\begin{acknowledgements}
  We would like to thank Meguru Komada for introducing us basic
  concept of Bayesian inference. YN was supported in part by JSPS
  KAKENHI Grant Number 15K05073 and 16H01094.
\end{acknowledgements}


\appendix
\section{Classical solution}
From the Hamiltonian \eqref{eq:Ht}, equations of motion for $\phi$
and $q$
are
\begin{align}
  &\frac{1}{N}\left(q^2\frac{\phi'}{N}\right)'+\mu^2q\phi=0,  \label{eq:eq1}\\
  &\frac{1}{N}\left(\frac{q'}{N}\right)'=-4q\left(\frac{\phi'}{N}\right)^2
    +2(1+\mu^2\phi^2),\\
  &\frac{1}{4}\left(\frac{q'}{N}\right)^2=q^2\left(\frac{\phi'}{N}\right)^2
    -1+q(1+\mu^2\phi^2). \label{eq:eq3}
\end{align}
By taking cosmic time $t$ as a time parameter and using the scale factor
$a=q^{1/2}$ and the original constants, \eqref{eq:eq1} and
\eqref{eq:eq3} become
\begin{align}
  &\ddot\Phi+3\left(\frac{\dot a}{a}\right)\dot\Phi+m^2\Phi=0,\\
  &\left(\frac{\dot
    a}{a}\right)^2+\frac{\Lambda}{3a^2}=\frac{\Lambda}{3}
    +\frac{4\pi G}{3}\left(\dot\Phi^2+m^2\Phi^2\right),
\end{align}
where~ $\dot{}=\frac{d}{dt}$.

Now we consider the solution of slow roll inflation driven by the mass
term in this model. The the slow roll condition is
\begin{equation}
  |\ddot\Phi|\lesssim\left(\frac{\dot a}{a}\right)|\dot\Phi|,\quad
  \dot\Phi^2\lesssim m^2\Phi^2,\quad \frac{\Lambda}{3}\lesssim\frac{4\pi
    G}{3}m^2\Phi^2, 
\end{equation}
and we also assume that the spatial curvature is negligible. Then the
scalar field evolves as
\begin{equation}
 \Phi\approx\Phi_\text{i}-\frac{m}{2\sqrt{3\pi G}}\,(t-t_\text{i}).  
\end{equation}
The domination of the mass term ends at $\Phi_\text{f}$,  which
depends on the value of a dimensionless parameter $\mu=m/\sqrt{\Lambda/3}$.
For $\mu<3$, 
\begin{equation}
  \Phi_\text{f}\approx\sqrt{\frac{\Lambda}{4\pi Gm^2}},\qquad
    \phi_\text{f}=\frac{1}{\mu},
\end{equation}
and below this value, the scalar field evolves as
\begin{equation}
  \Phi\approx\Phi_{\text{f}}\exp\left[-\frac{\mu^2}{3}\sqrt{\frac{\Lambda}{3}}
    \,(t-t_{\text{f}})\right].
\end{equation}
The e-foldings from $\Phi_{\text{i}}$ to $\Phi_{\text{f}}$ is
\begin{equation}
  \mathcal{N}=\ln\left(\frac{a_{\text{f}}}{a_{\text{i}}}\right)\approx\frac{3}{2}
 \left(\phi_{\text{i}}^2-\frac{1}{\mu^2}\right).
\end{equation}
For $\mu>3$,
\begin{equation}
  \Phi_{\text{f}}\approx \frac{1}{2\sqrt{3\pi
      G}},\qquad\phi_{\text{f}}=\frac{1}{3},
\end{equation}
and below this value, the scalar field oscillates around $\Phi=0$. The
e-foldings from $\Phi_{\text{i}}$ to $\Phi_{\text{f}}$ is
\begin{equation}
  \mathcal{N}=\ln\left(\frac{a_\text{f}}{a_\text{i}}\right)\approx 2\pi
  G(\Phi_\text{i}^2-\Phi_\text{f}^2)
  =\frac{3}{2}\left(\phi_\text{i}^2-\frac{1}{9}\right).
\end{equation}
The value of the scalar field at the Planck energy is defined
by\footnote{The Planck mass is defined by
  $m_{\text{pl}}^2=1/G$.}
\begin{equation}
  \frac{m^2}{2}\Phi^2_{\text{pl}}=m^4_\text{pl},\quad
  \phi_{\text{pl}}=\frac{4}{3}\frac{\sqrt{2K}}{\mu}.
\end{equation}
As $\phi_\text{f}<\phi_{\text{pl}}$, we have the following constraint
for parameters in our model
\begin{align}
  \mu<3:&\quad \frac{9}{32}<K,\\
  \mu>3:&\quad \frac{9}{32}<\frac{\mu^2}{32}<K.
\end{align}

\end{document}